\documentclass[a4paper,11pt]{article}
\pdfoutput=1

\usepackage{jinstpub}

\usepackage{lineno}

\usepackage[separate-uncertainty=true]{siunitx}
\DeclareSIUnit\au{AU}

\usepackage{amsmath}
\usepackage{xspace}
\usepackage[normalem]{ulem} 
\usepackage{booktabs}

\newcommand{\mycomment}[1]{}

\newcommand{\fig}[1]{figure~\ref{#1}}
\newcommand{\tab}[1]{table~\ref{#1}}

\newcommand{\secref}[1]{section~\ref{#1}}
\newcommand{\eqnref}[1]{eq.~\eqref{#1}}

\newcommand{\Fig}[1]{Figure~\ref{#1}}

\newcommand{\Secref}[1]{Section~\ref{#1}}
\newcommand{\Eqnref}[1]{Equation~\eqref{#1}}

\newcommand{\Nth}{\ensuremath{N^{\text{th}}}\xspace}
\newcommand{\ith}{\ensuremath{i^{\text{th}}}\xspace}
\newcommand{\jth}{\ensuremath{j^{\text{th}}}\xspace}
\newcommand{\Galphax}{\ensuremath{\mathcal{G}_{\alpha}(\vec{x})}\xspace}
\newcommand{\peXT}{\ensuremath{p_{\text{eXT}}}\xspace}
\newcommand{\peXTall}{\ensuremath{p_{\text{eXT},\text{all}}}\xspace}
\newcommand{\PeXTall}{\ensuremath{P_{\text{eXT},\text{all}}}\xspace}

\title{Characterization of a SiPM-based monolithic neutron scatter camera using dark counts}

\author[a,1]{J.~Balajthy,\begin{NoHyper}\note{Corresponding author.}\end{NoHyper}}
\author[c,d]{J.~Brown,}
\author[a]{E.~Brubaker,}
\author[a]{B.~Cabrera-Palmer,}
\author[c]{J.~Cates}
\author[c,d]{B.~L.~Goldblum,}
\author[c]{M.~Folsom,}
\author[b]{P.~Hausladen,}
\author[e]{K.~Keefe,}
\author[b]{J.~Nattress,}
\author[c]{V.~Negut,}
\author[e]{K.~Nishimura,}
\author[a]{J.~Steele,}
\author[b]{and K.~Ziock}

\affiliation[a]{Sandia National Laboratories, Livermore, CA}
\affiliation[b]{Oak Ridge National Laboratory, Oak Ridge, TN}
\affiliation[c]{Lawrence Berkeley National Laboratory, Berkeley, CA}
\affiliation[d]{University of California at Berkeley, Berkeley, CA}
\affiliation[e]{University of Hawai'i at  M\={a}noa, Honolulu, HI}

\emailAdd{jabalaj@sandia.gov}

\collaboration[c]{on behalf of the SVSC collaboration}

\abstract{The Single Volume Scatter Camera (SVSC) Collaboration aims to develop portable neutron imaging systems for a variety of applications in nuclear non-proliferation.
Conventional double-scatter neutron imagers are composed of several separate detector volumes organized in at least two planes. A neutron must scatter in two of these detector volumes 
for its initial trajectory to be reconstructed. As such, these systems typically have a large footprint and poor geometric efficiency. We report on the design and characterization of a prototype 
monolithic neutron scatter camera that is intended to significantly improve upon the geometrical shortcomings of conventional neutron cameras.
The detector consists of a $\SI{50}{mm}\times\SI{56}{mm}\times\SI{60}{mm}$ monolithic block of EJ-204 plastic scintillator instrumented on two faces with arrays of 64 Hamamatsu S13360-6075PE silicon photomultipliers (SiPMs).
The electronic crosstalk is limited to \SI{< 5}{\percent} between adjacent channels and \SI{< 0.1}{\percent}
between all other channel pairs. 
SiPMs introduce a significantly elevated dark count rate over PMTs, as well as correlated noise from after-pulsing and optical crosstalk. In this article, we characterize the dark count rate 
and optical crosstalk and present a modified event reconstruction likelihood function that accounts for them. We find that the average dark count rate per SiPM is \SI{4.3}{MHz} with a standard deviation of \SI{1.5}{MHz} among devices. 
The analysis method we employ to measure internal optical crosstalk also naturally yields the mean and width of the single-electron pulse height. We calculate separate contributions to the width 
of the single-electron pulse-height from electronic noise and avalanche fluctuations. We demonstrate a timing resolution for a single-photon pulse to be \SI[separate-uncertainty = true]{128 (4)}{ps}. 
Finally, coincidence analysis is employed to measure external (pixel-to-pixel) optical crosstalk. We present a map of the average external crosstalk probability between $2\times4$ groups of SiPMs, as well as the in-situ timing characteristics extracted from the coincidence analysis.
Further work is needed to characterize the performance of the camera at reconstructing single- and double-site interactions, as well as image reconstruction.
}

\keywords{Neutron detectors (cold, thermal, fast neutrons), Search for radioactive and fissile materials, Inspection with neutrons, Analysis and statistical methods}

\begin{document}
\maketitle
\flushbottom

\section{Introduction}
\label{sec:intro}
The Single Volume Scatter Camera (SVSC) Collaboration aims to develop a novel type of kinematic neutron imager that utilizes a single, compact detector volume for the purpose of imaging fission-energy neutrons (\SIrange{1}{10}{MeV}). Conventional double-scatter neutron imagers rely on several separate detectors arranged in at least two detection planes~\cite{Vanier2007,Bravar2009,Brennan2011,Goldsmith2016}. The SVSC collaboration is pursuing cameras with several different compact form factors, including two prototypes that utilize a single, monolithic scintillating block as a detector volume that is instrumented on at least two sides with pixelated photon detectors~\cite{braverman2018,keefe2022}. Previous modeling studies have suggested that such a detector may have as much as an order of magnitude increase in efficiency for neutron double-scatter detection over a traditional kinematic neutron imager~\cite{braverman2018}. It would also benefit from reduced size and weight, which would improve deployability. A smaller, compact detector would also be able to be located closer to a source, maximizing the solid angle efficiency of the system and improving spatial resolution for a given angular resolution. 

\begin{figure}
\centering 
\includegraphics[width=.7\textwidth]{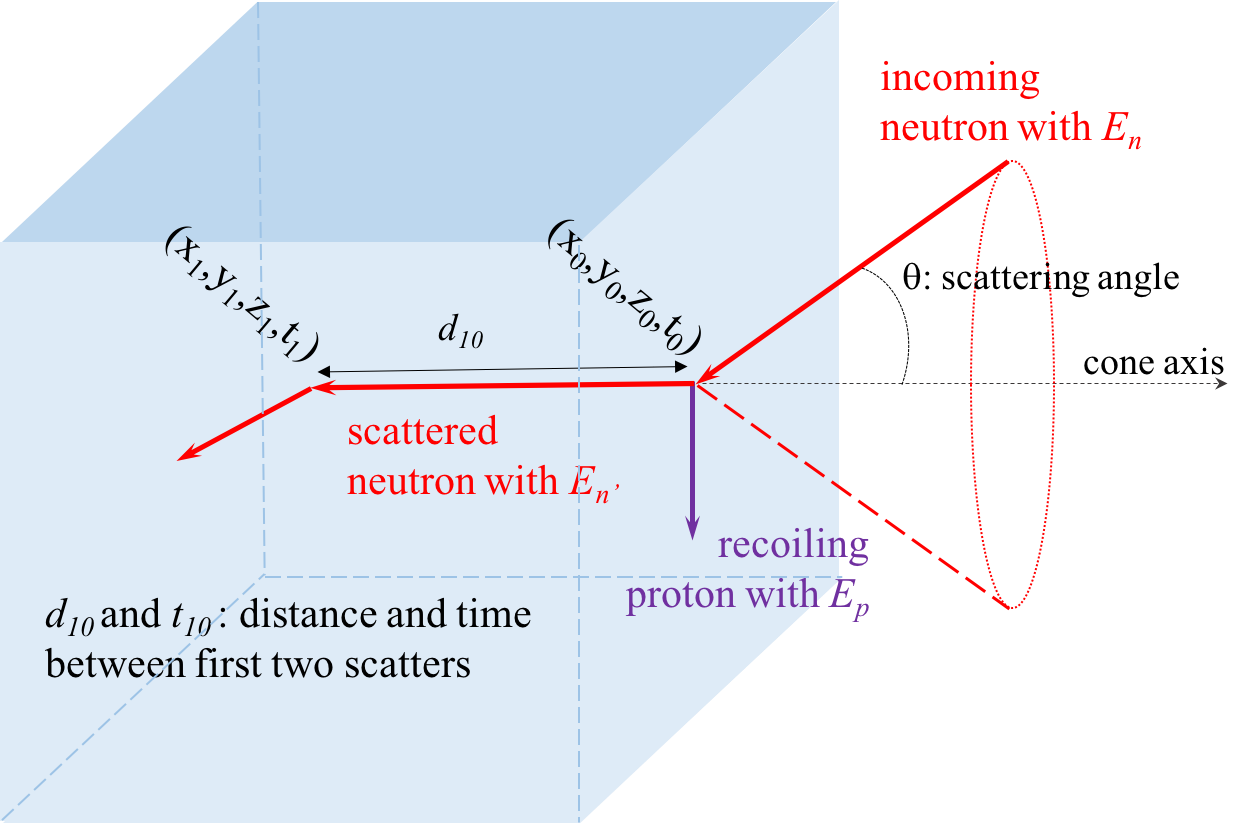}
\caption{\label{fig:eventdiagram} Diagram of a double-scatter event in a neutron kinematic imaging detector \cite{braverman2018}.}
\end{figure}

The technique of imaging a neutron source using a scatter camera relies on high-precision measurement of the incoming neutron energy ($E_n$), the energy after the first interaction ($E_n'$), and the positions of the two scattering locations. $E_n'$ is determined by the neutron time-of-flight equation which necessitates measurement of the scattering times as well:
\begin{equation}
E_n' = \frac{1}{2}m_n \left ( \frac{\Delta d}{\Delta t} \right )^2.
\end{equation}
The total neutron energy, $E_n$, is calculated by adding $E_n'$ to the energy deposited in the first scatter location as measured by the detector response. The positions of the two interactions provide the endpoints of the trajectory of the neutron between the first and second interaction. The scattering angle, $\theta$, of the first interaction can be determined from $E_n$ and $E_n'$:
\begin{equation}
\cos (\theta) = \sqrt{\frac{E_n'}{E_n}}.
\end{equation}
The incident trajectory of the neutron can therefore be reconstructed up to a conical degeneracy as shown in \fig{fig:eventdiagram}.

The monolithic neutron scatter camera described in this document is designed so that the first two interactions occur in a single, unsegmented detector volume. The photons from these interactions are collected by pixelated photon detectors that are coupled to the exterior surface of the detector volume. The positions and times of these photon detections, along with the calibrated detector response, is used to reconstruct the positions, times, and energies of the two interactions. Such a detector based on a \SI{10}{cm} cube of EJ-232Q has been shown via modeling to have a predicted imaging resolution of \SI{0.101}{rad} \cite{ej232q_datasheet,braverman2018}.

The analysis chain for this detector can be divided into three steps: single-photon processing, event reconstruction, and image reconstruction. The detector is designed so that each photodetector pixel is capable of providing a list of arrival times for each photon detected during an event. The interaction-level information can then be determined by performing a maximum likelihood analysis on the pixel positions and arrival times of the detected photons. Finally, image reconstruction can be performed by employing a conventional algorithm such as back-projection or maximum likelihood expectation maximization on the interaction-level information.

In \secref{sec:detector}, we describe the design and commissioning of the detector, including the design of the electronic readout system, detector materials, and assembly. In \secref{sec:dcrintro}, we describe SiPM-specific noise and a method of measuring SiPM characteristics in situ using dark count data. \Secref{sec:likelihood} describes a modified likelihood model for event reconstruction that includes terms for SiPM-specific noise and enumerates the parameters that need to be measured to employ this model. \Secref{sec:sptr} describes our single-photon processing method and its performance as characterized by dark count data and measurements with a fast pulsed laser. \Secref{sec:darkcounts} describes an analysis of the pulse-height spectrum measured from dark count data, and \secref{sec:timecorr} describes a time correlation analysis of dark count data that allows us to characterize the timing performance of the SiPMs in situ as well as the external optical crosstalk. 

\section{Detector Design}
\label{sec:detector}
\begin{figure}
\centering 
\includegraphics[width=.5\textwidth]{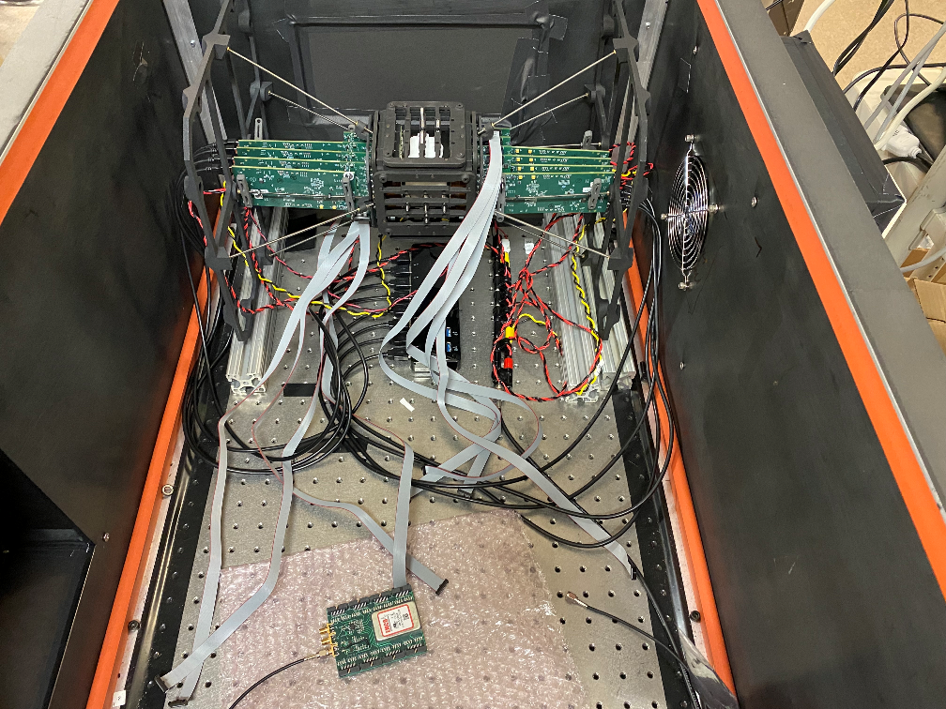}
\qquad
\includegraphics[width=.4\textwidth]{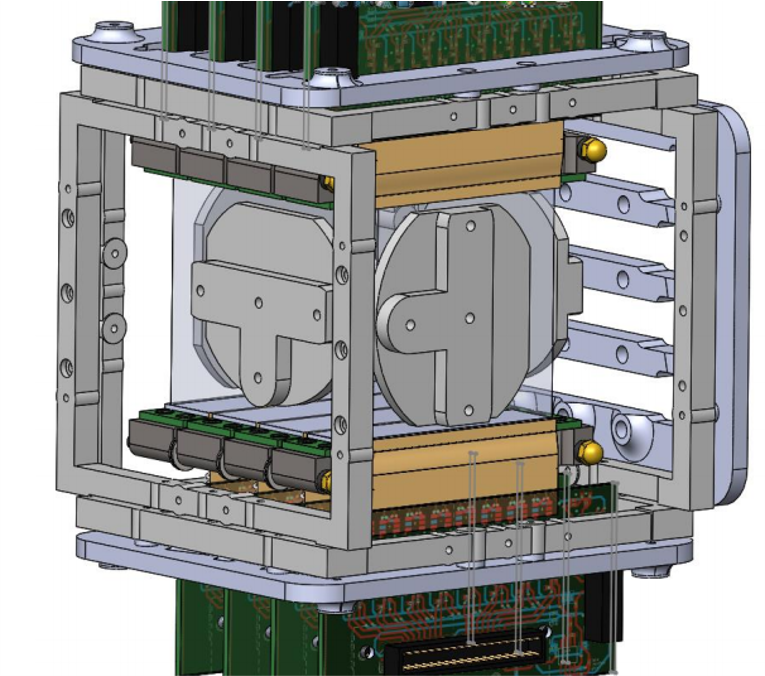}
\caption{\label{fig:assembly} (Left) Picture of the assembled detector inside the dark box. (Right) CAD drawing of the detector mount. The scintillator (shown as the semi-transparent region) is held in place using spring-loaded plastic plates on the four non-instrumented sides. The SiPM arrays are compressed against opposite sides of the scintillator also using springs.}
\end{figure}

\begin{figure}
\centering 
\includegraphics[width=.5\textwidth]{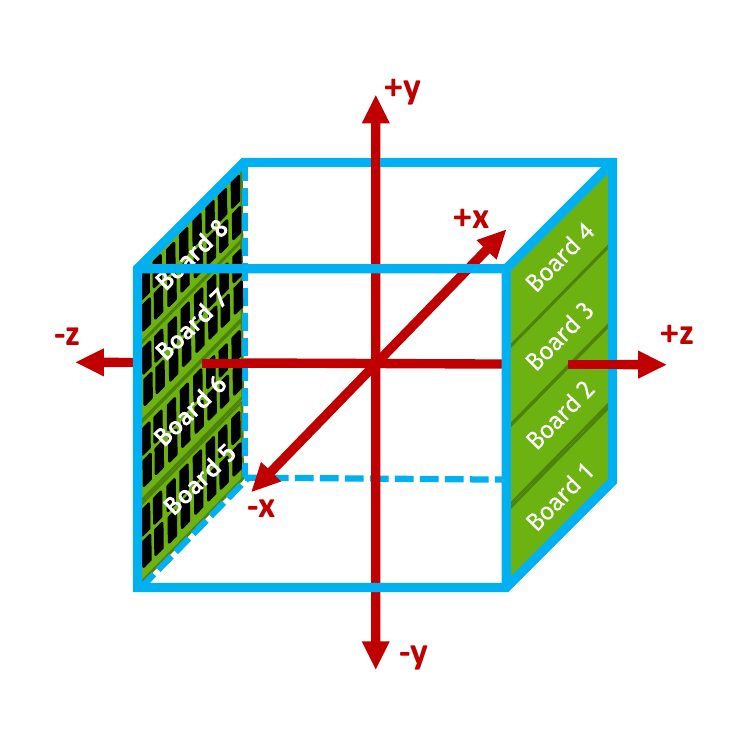}
\caption{\label{fig:monocartoon} Orientation of detector assembly. The long axis of the $2 \times 8$ SiPM arrays are aligned parallel to the $x$ axis, and the short axis is parallel to the $y$ axis. The arrays are aligned perpendicular to the $z$ axis. The origin is defined to be the center of the detector. The ``Board'' numbering scheme shown in the figure applies to both the front-end and SCEMA boards.
}
\end{figure}

\begin{figure}
\centering 
\includegraphics[width=.99\textwidth]{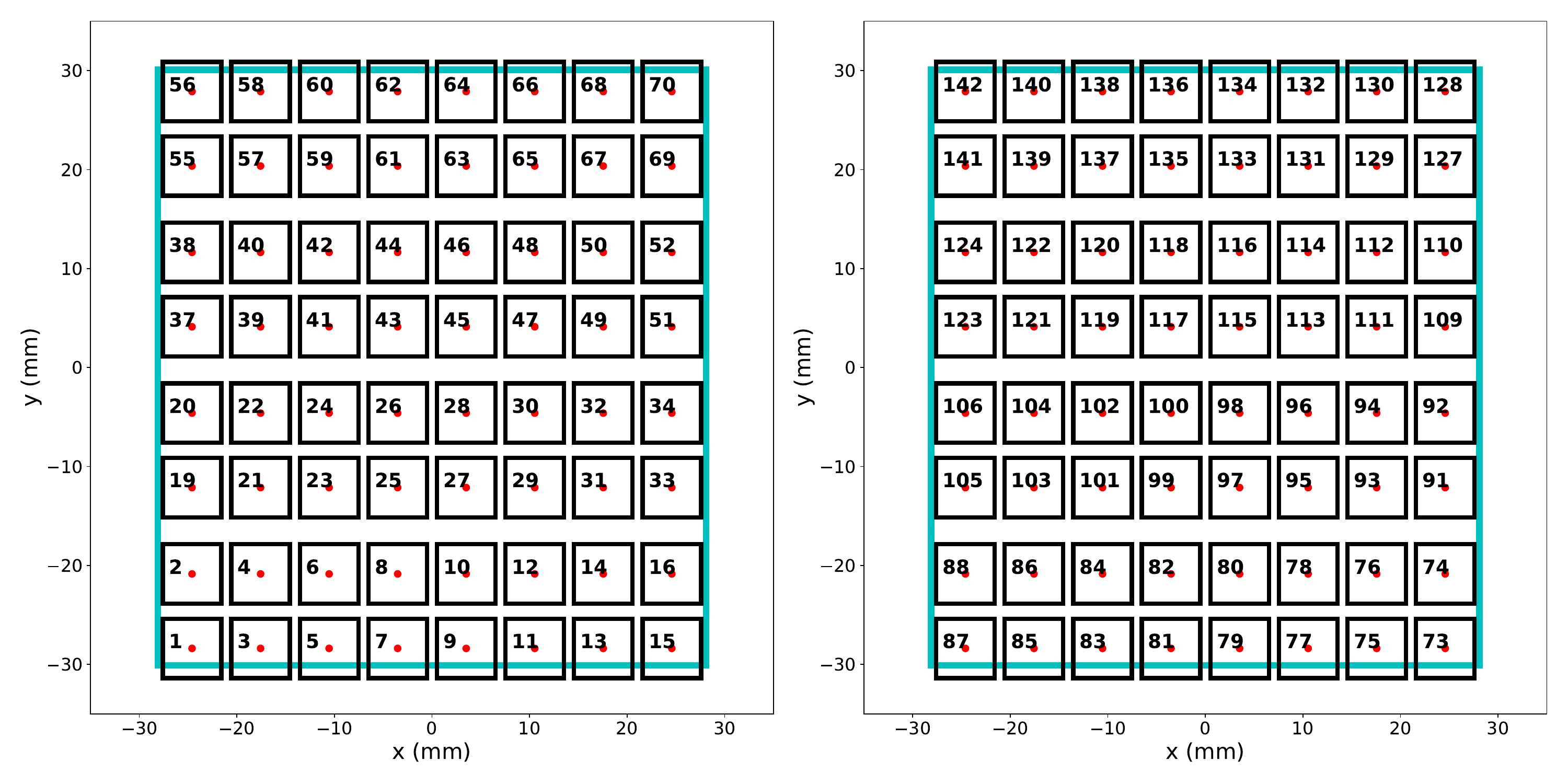}
\caption{\label{fig:channellayout} Layout of SiPM channels for each 8x8 array. The black rectangles indicate the edges of the sensitive area of each SiPM, and the red dots indicate the center points. The
 numbers indicate the SCEMA channel number that is reading out each SiPM. The large cyan rectangle indicates the edge of the scintillator block. The edges of the scintillator are located at
 $x=\SI{\pm 28.1}{mm}$, $y=\SI{\pm 30.1}{mm}$, $z=\SI{\pm 25}{mm}$. The left hand panel represents the array located at $z=\SI[retain-explicit-plus]{+25}{mm}$, and the right hand plot
 represents the array located at $z=\SI{-25}{mm}$. }
\end{figure}

The prototype uses a ($\SI{50}{mm}\times\SI{56.2}{mm}\times\SI{60.2}{mm}$) block of EJ-204 plastic scintillator from Eljen as the detection medium~\cite{ej204_datasheet}. 
The EJ-204 block is instrumented on two sides with Hamamatsu S13360-6075PE multi-pixel photon counters (MPPCs) as described in
figures~\ref{fig:assembly},~\ref{fig:monocartoon},~and~\ref{fig:channellayout}~\cite{s13360_datasheet}. The Hamamatsu MPPCs are two-dimensional arrays of Geiger mode avalanche photodiodes 
(G-APDs) and are more commonly referred to as silicon photomultipliers (SiPMs)~\cite{acerbi2019}. The SiPMs are mounted onto eight custom-built front-end circuit boards, each of which houses a $2 \times 8$ array of SiPMs \cite{cates2022}. Four of the $2 \times 8$ arrays are mounted to each of two opposite sides of the EJ-204 block so that each side is instrumented by an $8 \times 8$ array of SiPMs as is shown in \fig{fig:monocartoon}. The specific arrangement of the SiPMs is shown in \fig{fig:channellayout}. There is an additional \SI{1.2}{mm} gap between each of the $2 \times 8$ arrays over what was accounted for in the initial design. These added gaps cause the top and bottom row of SiPMs to slightly overhang the scintillator. In total, the active area of the SiPMs covers \SI{63}{\percent} of the instrumented faces. In order to reduce reflections, the remaining four non-instrumented sides of the scintillator have been sanded with 400 grit sandpaper and painted with EJ-510B black latex paint~\cite{ej510_datasheet}. For the event reconstruction analysis described in \secref{sec:likelihood} we are only interested in photons that travel directly from a scintillation event site to the SiPM in which it is detected. Any photons that are reflected prior to detection are not treated by the analysis and will introduce bias and uncertainty due to model mismatch. 

The SiPM arrays are coupled to the scintillator via EJ-560 optical interface pads~\cite{ej560_datasheet}. A custom 3D printed mechanical structure (shown in the right panel of \fig{fig:assembly}) compresses each $2 \times 8$ SiPM array against the optical interface using three springs. The detector assembly is housed inside of a dark box, as shown in the left panel of \fig{fig:assembly}. The dark box is cooled directly using a portable air conditioning (AC) unit~\cite{trippliteAC_datasheet}. The cold air output of the AC unit is piped into the dark box through a light-tight louver, and a cooling fan constantly exhausts air from the box through another light-tight louver on the opposite side.

\begin{figure}
\centering 
\includegraphics[width=.9\textwidth]{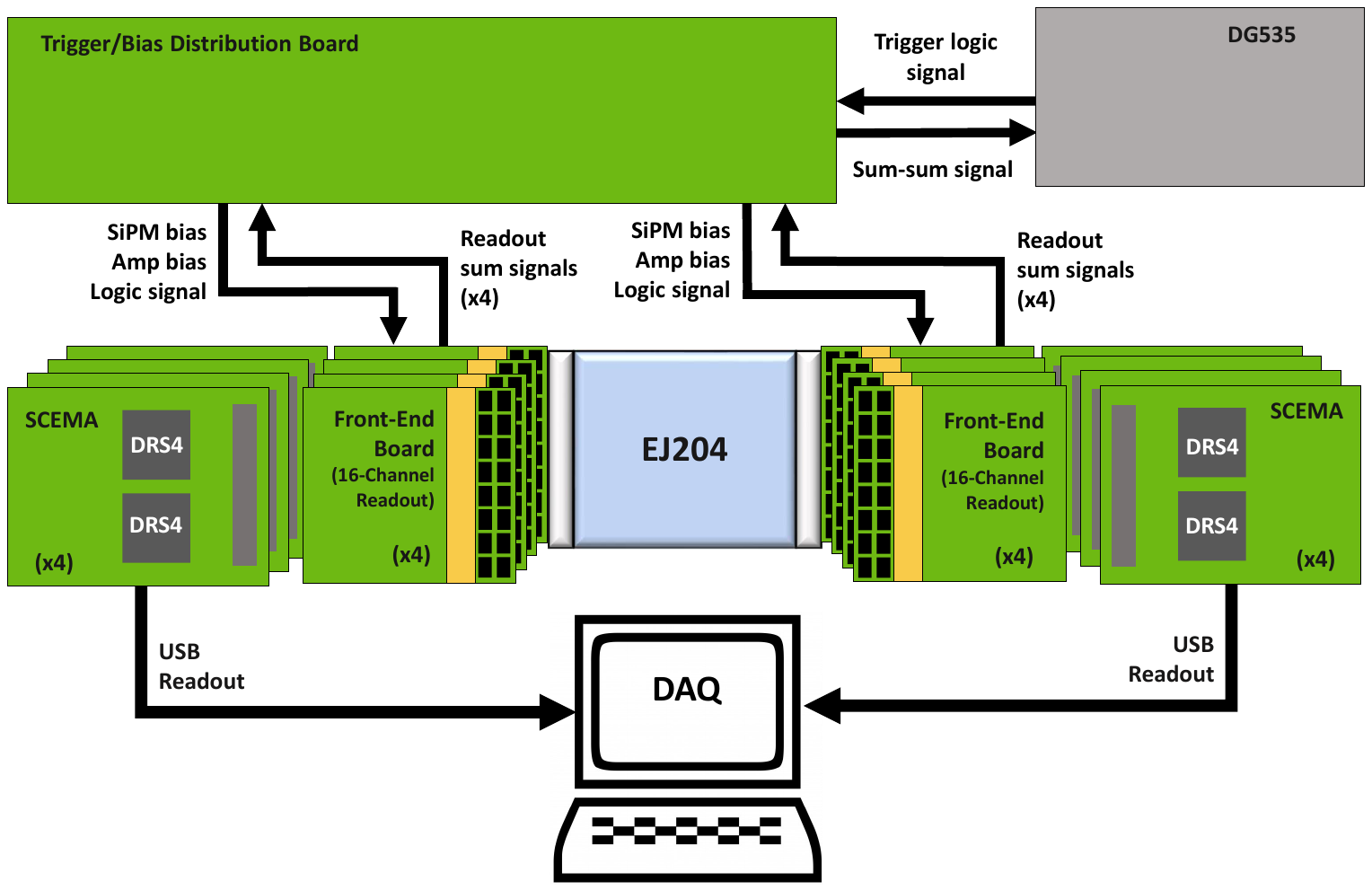}
\caption{\label{fig:readout} Readout scheme for the detector as described in the text. A block of EJ-204 scintillator is instrumented on two sides with an $8 \times 8$ array of SiPMs, coupled using EJ-560 optical interface pads \cite{ej204_datasheet,ej560_datasheet}. Each $8 \times 8$ array is composed of four $2 \times 8$ arrays that are housed on a custom front-end board that provides bias, readout, amplification, and signal summing \cite{cates2022}. A trigger/bias distribution board provides bias to the 128~SiPMs and RF amplifiers. It also handles trigger distribution to and from the front-end boards. An SRS DG535 function generator provides a common logic signal that can be either random or based on the sum-sum signal from the trigger/bias distribution board~\cite{dg535_datasheet}. Each front-end board passes its 16~amplified SiPM signals, along with the sum and trigger signal, to a SCEMA electronics board for digitization. The SCEMAs are controlled and read out via USB connection to a data acquisition computer. }
\end{figure}

\begin{figure}
\centering 
\includegraphics[width=.4\textwidth]{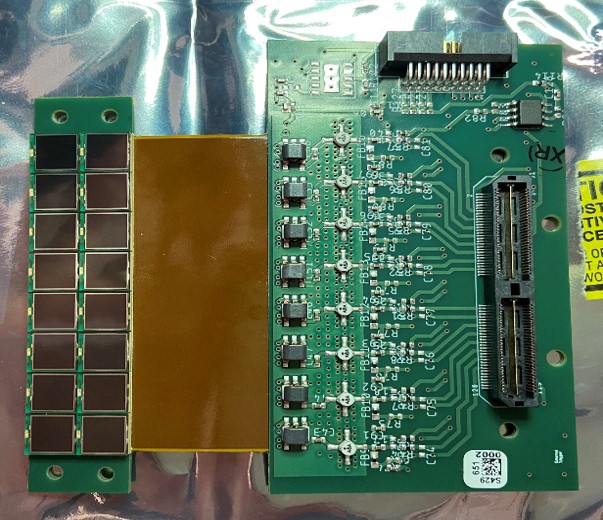}
\includegraphics[width=.55\textwidth]{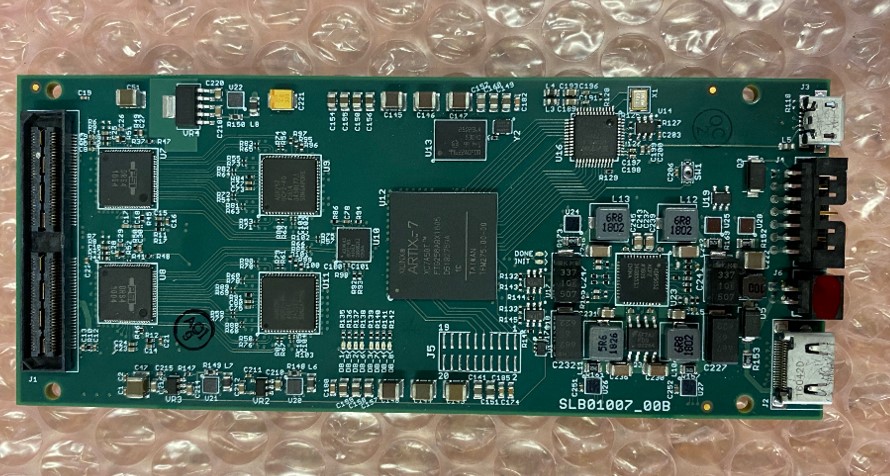}
\caption{\label{fig:frontend} (Left) Custom-designed SiPM front-end board. (Right) SCEMA-b waveform capture digitization board.}
\end{figure}

The readout scheme for the SiPMs is summarized in \fig{fig:readout}. 
The SiPM front-end boards (shown in the left-hand image of \fig{fig:frontend}) house RF amplifiers for each channel to allow
single-photon sensitivity and a summing circuit to provide a trigger signal. They are designed to have minimal electronic crosstalk while also maintaining a high geometric efficiency for photon detection. The device selection and readout board development is described in~\cite{cates2022}.

For triggering, each front-end board includes a summing circuit that combines the signals from each of the 16 SiPM channels. The sum signals from all the front-end boards are passed to a custom trigger/bias distribution board, where they are themselves summed together into a sum-sum signal. The sum-sum signal is itself passed to a SRS model DG535 pulse generator~\cite{dg535_datasheet}, which looks for a rising-edge threshold crossing. The output logic signal from the DG535 is passed back to the trigger/bias distribution board. The DG535 is also able to provide random triggering that does not depend on the sum-sum signal. The trigger/bias distribution board houses voltage regulators that provide bias for all 128~SiPMs and RF amplifiers. The trigger/bias distribution board passes the SiPM bias, RF amplifier bias, and the common logic signal from the DG535 to the eight front-end boards. The trigger/bias distribution board provides a \SI{9.2}{V} bias to the RF amplifiers and a \SI{56}{V} bias to all 128 SiPMs. The breakdown voltages for the SiPMs range from \SI{51.16}{V} to \SI{52.61}{V}, so the over-voltage for each pixel is not precisely known. The over-voltage affects several SiPM parameters including gain, dark rate, crosstalk probability, and photon detection efficiency. These parameters must be quantified in order to perform event reconstruction and so must be measured using data.

Each of the eight front-end boards passes its 16~amplified SiPM channel readouts, its sum signal, and the common logic signal from the DG535 to an associated version b Sandia Laboratories Compact Electronics for Modular Acquisition (SCEMA) electronics board~\cite{steele2019,keefe2022} (right-hand image in \fig{fig:frontend}) for digitization via domino ring sampler (DRS4) switched capacitor array chips~\cite{ritt2010} at about \SI{5}{GS/s} sampling rate. The common logic signal is also passed to a comparator on each of the SCEMAs that initiates the digitization and data transfer. The rising edge of the digitized common logic signal is used to align the timing of the traces. The SCEMA (version a) performance is documented in ~\cite{steele2019}; version b is designed to allow more flexibility in the front end, but has similar design and performance as concerns the core waveform capture functionality~\cite{keefe2022}.  A back-end hub board was designed to synchronize the SCEMA clocks, perform more advanced trigger logic, and enable higher rate data transfers, but is not implemented for the results reported here. 

When a SCEMA is triggered, the control software application initiates the data transfer via USB. After all eight SCEMAs are triggered and complete their data transfer, the control application rearms them sequentially. The total time it takes for all eight SCEMAs to perform the digitization and data transfer for a single event is expected to be less than \SI{5}{ms}. In the current system without a back-end hub board, the actual event acquisition time is about \SIrange{100}{200}{ms}, dominated by the software execution time. This gives a maximum data acquisition rate of about \SI{10}{Hz}. The sequential arming of the SCEMAs allows for trigger mismatch, caused by early SCEMAs in the arming chain being triggered before the later SCEMAs are armed. We implement a \SI{300}{ms} hold-off on the DG535 after each event in order to prevent this mismatch. This data acquisition rate is acceptable for the analysis performed here, as we only require a single high-statistics data-set consisting of 130,000 trigger windows, which took three days to acquire. A full calibration data-set would require many times that number of trigger windows and would take months to acquire. Similarly, when the detector is imaging a neutron source such as AmBe or Cf-252 there will typically be many more triggers on gamma events and single site-neutron events than on double-site neutron events, so a single image may take days or weeks to complete, absent the improvements enabled by the back-end hub board.

We measure the electronic crosstalk using a Photek~LPG-405 pulsed laser~\cite{lpg405} incident on an area
spanning two SiPMs. Using two SiPMs instead of one allows us to increase the total intensity incident on the array without saturating a SiPM
and also allows us to test channels from both DRS4 chips on a SCEMA at the same time. The laser was attenuated using neutral density filters 
until the amplitude of the response in the target pixels was about \SI{150}{mV}. The time-correlated amplitudes in each of
 the 16~channels of a front-end board are measured relative to the amplitudes in the illuminated cells. We find the electronic crosstalk to 
be \SI{< 5}{\percent} between adjacent channels and 
\SI{< 0.1}{\percent} between all other channel pairs. This level of electronic crosstalk is well below the peak-finding threshold for single-photon counting analysis.

\section{Dark Count Spectrum Analysis}
\label{sec:dcrintro}
Noise counts in SiPMs occur when an avalanche is triggered by something other than a primary photon and may or may not be correlated to a previous avalanche. These noise counts fall into three categories: dark counts (uncorrelated), after-pulses (correlated), and optical crosstalk (correlated)~\cite{acerbi2019}. The dark count rate (DCR) in SiPMs is high enough to be non-negligible; for an event lasting \SI{18}{ns} (10 times the main component of the scintillator decay time), the 128 S13360-6075PE SiPMs required for two-sided instrumentation are expected to produce up to 15 dark counts~\cite{s13360_datasheet}. A \SI{244}{keV} neutron scatter, which is close to the expected lower limit of our region of interest, is expected to yield about 224 scintillation photons in EJ-204~\cite{ej204_datasheet,Laplace2020}. For an event at the center of the detector, about 28 of these photons would be detected given the photon detection efficiency and solid angle subtended by the SiPMs. Dark counts are an uncorrelated form of noise, so the impact they have on reconstruction resolution is constant in magnitude. Higher energy events would be less impacted by dark counts. Additionally, the dark counts would be distributed uniformly across the entire \SI{18}{ns} event window, whereas the signal photons would be concentrated around the interaction times.

The impacts of correlated noise counts are not mitigated by timing and signal size in the way that uncorrelated counts are. For that reason, the correlated noise incidence is likely to place a greater limit on the detector resolution. After-pulsing occurs when a charge carrier from a previous avalanche becomes trapped in a non-sensitive region of the SiPM before drifting back and causing a secondary avalanche in the same G-APD. Every avalanche in a G-APD emits a number of secondary photons which can trigger secondary avalanches in a process referred to as optical crosstalk. Optical crosstalk events can occur in the same SiPM as the initial avalanche (internal crosstalk, iXT), or in a different device if the secondary photon escapes the original SiPM (external crosstalk, eXT). Well-matched indices of refraction at the optical interface may reduce internal reflections, allowing some crosstalk photons to escape that otherwise would have been reflected back into the primary SiPM, thereby increasing eXT and decreasing iXT~\cite{Masuda2021}. 

According to the S13360 datasheet, the probability of after-pulses is expected to be small, and the correlated noise is expected to be dominated by optical crosstalk~\cite{s13360_datasheet}. The S13360 datasheet quotes a single value for the optical crosstalk probability, which is between \SI{10}{\percent} and~\SI{20}{\percent} depending on the operating voltage. The crosstalk probability as defined in the datasheet corresponds only to iXT and may differ from the measured iXT depending on the properties of the optical interface.

The significant dark count rate and incidence of crosstalk makes dark count measurements a useful calibration tool. The pulse-height spectrum from a set of data acquired with a random trigger can be analyzed to calculate the dark count rate, crosstalk probability, SiPM gain, gain fluctuation, and electronic noise.  A dark count is caused by thermal emission of a single electron in a G-APD, which causes an avalanche in that G-APD. This may then cause subsequent G-APDs to trigger via optical crosstalk. This means the pulse-height spectrum of SiPM dark counts displays peaks at integer multiples of the mean single-electron pulse-height. The relative size of the peaks is determined by optical crosstalk probability. The probability distribution of the number of crosstalk avalanches can be approximated by a Borel distribution up to a multiplicity of about five~\cite{biland2014}. A Borel distribution describes a branching Poisson process. It is the probability to have exactly $N$ breakdowns assuming that each breakdown produces a Poisson number of additional breakdowns with mean~$\mu$. The probability mass function is given by:
\begin{equation}
P_{\text{Borel}}(N,\mu)=\frac{(\mu N )^{N-1}}{N(N-1)!}e^{-\mu N}
\end{equation}
This distribution is attractive because it is physically motivated and because it is simple to implement numerically. The Borel distribution does not take into account the fact that there are fewer nearby untriggered G-APDs available at high multiplicities and so overestimates the probability of crosstalk above about $N=5$.

The voltage separation of the peaks in the pulse-height spectrum is a direct measure of the SiPM gain ($x_g$). The width of the \Nth multiplicity peak ($\sigma_N$) has contributions from the gain fluctuations ($\sigma_a$) and from electronic noise ($\sigma_e$):
\begin{equation}
\sigma_N^2 = \sigma_e^2+N\sigma_a^2.
\end{equation}
The \Nth multiplicity peak can be fit to the following Gaussian:
\begin{equation}\label{eq:ser_gausfit}
Y_N = \frac{A\,P_{\text{Borel}}(N,\mu) }{\sqrt{2 \pi} \sigma_N}e^{-(x-Nx_g-x_0)^2/(2\sigma_N^2)},
\end{equation}
where the fitting parameter~$A$ is equal to the total number of counts summed over all multiplicity peaks. The fitting parameter~$x_{0}$ allows for a non-zero offset in the SiPM response and is typically found to be very small. 

\begin{figure}
\centering 
\includegraphics[width=.6\textwidth]{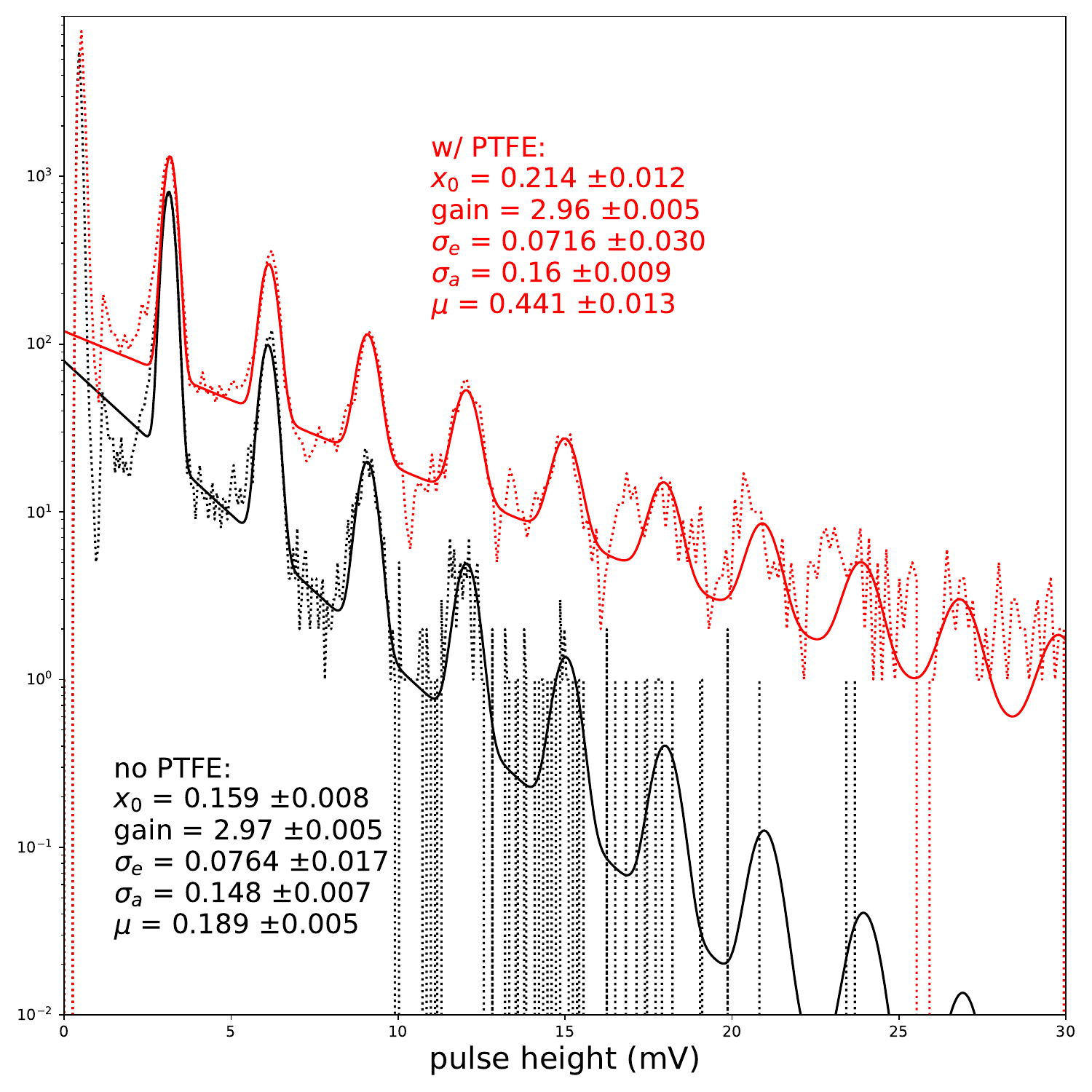}
\caption{\label{fig:dcprelim} Data from a single S13360-6075PE SiPM. The fitting function is described in the text. The dotted red lines show data from when the SiPM was wrapped in Teflon, and the dotted black lines show data from the unwrapped SiPM. The solid red and black likes show the best-fit models for the wrapped and unwrapped data, respectively. }
\end{figure}

\Fig{fig:dcprelim} shows the model described in \eqnref{eq:ser_gausfit} fit to a set of data taken using a single SiPM mounted on a test stand and biased at \SI{58}{V}~\cite{cates2022}. The fitting model also includes an exponentially falling baseline. We believe the counts contained in this apparent exponential baseline are true SiPM counts that have been smeared by systematic effects such as after-pulsing and pile-up. Our fitting method would be improved by including a more physical treatment of these counts.

The dark count data was taken both with (red lines) and without (black lines) a Teflon wrapping around the SiPM. The Teflon wrapping will reflect any eXT photons back to the SiPM, so the difference between the measured crosstalk parameters from the two fits gives the approximate value of the eXT crosstalk parameter. 
The results indicate an iXT parameter of \num{0.189 \pm 0.005} and an eXT parameter of \num{0.252 \pm 0.014}, which are in line with the datasheet value. The dark count rate increases from about \SI{2.3}{MHz} with no Teflon wrapping, to \SI{3.2}{MHz} with Teflon wrapping. It is not clear why the rate increases when the SiPM is wrapped in Teflon. It may be due to increased temperature or some other physical interference that causes increased thermal emission. We also expect some increase in dark count rate from increased after-pulsing and delayed optical crosstalk, however these effects should be much smaller than what is observed. 

The unexplained rise in DCR in the Teflon-wrapped SiPMs demonstrates that SiPM noise parameters can vary significantly depending on the specific setup of the assembled detector. The likelihood model described in~\secref{sec:likelihood} requires accurate values of the noise parameters, including the DCR, in order to reconstruct unbiased results. It is therefore important to be able to characterize the in-situ noise parameters of an assembled detector as is done in \secref{sec:darkcounts} and \secref{sec:timecorr}. For this reason, the SiPM parameters measured in this section will be used only as points of comparison for these in-situ measurements.

\section{Likelihood Analysis of Interations}\label{sec:likelihood}
In~\cite{braverman2018}, the authors propose the use of an unbinned maximum likelihood technique for event reconstruction in a monolithic scatter camera. The likelihood model they use is idealized and does not account for SiPM-specific systematic effects such as high DCR and optical crosstalk. In this section, we use the likelihood model of~\cite{braverman2018} as a starting point and describe a set of modifications that allows for unbiased reconstruction of event information in the presence of SiPM-specific noise.

The interaction level information described in \secref{sec:intro} can be extracted from the list of detected photons by maximizing the extended likelihood,~$\mathcal{L}$~\cite{Barlow_1990,braverman2018}:
\begin{equation}\label{eq:extlik}
\mathcal{L} = \frac{\nu^n}{n!}e^{-\nu}\prod_{i=1}^{n} p_i(x_1,...,x_m;a_1,...,a_k),
\end{equation}
where $\nu$ is the total number of expected photons, $n$ is the number of detected photons, $(x_1,...,x_m)$ is a set of observable quantities, and $(a_1,...,a_k)$ is a set of fitting parameters. In \eqnref{eq:extlik}, $p_i(x_1,...,x_m;a_1,...,a_k)$ is the normalized probability of observing  $(x_1,...,x_m)$ given the set of trial parameters $(a_1,...,a_k)$. The extended likelihood allows the expected number of samples (e.g. photons) in a data-set, $\nu$, to be a variable in the fit. The usual normalization condition on the probability distribution applies:
\begin{equation}\label{eq:liknorm}
\int...\int p_i(x_1,...,x_m;a_1,...,a_k)dx_1 ... dx_m  = 1
\end{equation}

\Eqnref{eq:maxlik_ideal} describes the ideal version of the likelihood for a given set of photons detected from an event with $N$ interaction sites:
\begin{equation}\label{eq:maxlik_ideal}
\mathcal{L}=\frac{\nu^n}{n!}e^{-\nu}\prod_{i=1}^{n} \ \frac{1}{\nu} \ \sum_{j=1}^{N}\mathcal{N}_j \ \mathcal{G}_{\alpha(i)}(\vec{x}_j) \ \mathcal{F}_{\alpha(i)}(t_i-t_j-| \vec{x}_j-\vec{x}_{\alpha(i)} |/c_p).
\end{equation}
In this likelihood definition, the detected photons are indexed by $i$. The detector pixels are indexed by $\alpha$, and the pixel that detects the $i_{th}$ photon is denoted by $\alpha(i)$. The location of the center of the face of pixel $\alpha$ is given by $\vec{x}_{\alpha}$. The proposed interactions are indexed by $j$. The $j^{th}$ interaction occurs at location $\vec{x}_j$ at time $t_j$ and emits $\mathcal{N}_j$ photons. The predicted energy deposited at the $j^{th}$ interaction site is given by:
\begin{equation}
\langle E \rangle_j = \frac{\mathcal{N}_j}{ Y_{l} },
\end{equation}
where $Y_l$ is the absolute light yield of the scintillator. The geometric efficiency, \Galphax, is the probability that a photon generated at position $\vec{x}$ will be detected by the pixel detector with index $\alpha$. In absence of reflections and other confounding effects, the geometric efficiency 
is given by:
\begin{equation}\label{eq:eff_ideal}
\Galphax = \frac{\Omega_{\alpha}(\vec{x})}{4\pi}\mathcal{E}_{\alpha} \ e^{-| \vec{x}-\vec{x}_{\alpha} |/\lambda},
\end{equation}
where $\Omega_{\alpha}$ is the solid angle subtended by pixel~$\alpha$ from the interaction site, $\mathcal{E}_{\alpha}$ is the photon detection efficiency of pixel~$\alpha$, and $\lambda$ is the attenuation length of scintillation light in the medium. The true number of detected photons in \eqnref{eq:maxlik_ideal} is $n$, and the total expected number of detected photons is $\nu$. Each of the $N$ interaction sites contributes an expected $\mathcal{N}_j  \Gamma_j$ photons, where:
\begin{equation}
\Gamma_j = \sum_{\alpha = 1}^{N_{\text{SiPM}}}\mathcal{G}_{\alpha}(\vec{x_j})
\end{equation}
is the total probability that a photon from the $j^{th}$ interaction site will be detected in any pixel. The time distribution,~$\mathcal{F}_{\alpha}(t_i-t_j-| \vec{x}_j-\vec{x}_{\alpha(i)}|/c_p)$, is the scintillation time response convolved with time response of pixel $\alpha(i)$. 
The time distribution is delayed by the photon travel from the interaction site to the pixel, $| \vec{x}_j-\vec{x}_{\alpha(i)} |/c_p$, where $c_p$ is the speed of light in the medium. 
When integrated over time and summed over all available pixels, each interaction term has a weight equal to the expected number of photons detected from that interaction, $\nu_j$:
\begin{equation}
\int\sum_{\alpha} \mathcal{N}_j \ \mathcal{G}_{\alpha(i)}(\vec{x}_j) \ \mathcal{F}_{\alpha}(t_i-t_j-| \vec{x}_j-\vec{x}_{\alpha(i)} |/c_p) dt \ = \mathcal{N}_j \cdot \Gamma_j \cdot 1= \nu_j,
\end{equation}
so in order to satisfy the normalization condition in \eqnref{eq:liknorm}, the sum is divided by~$\nu = \sum_j \nu_j$.

\Eqnref{eq:maxlik_ideal} does not account for the noise counts described in \secref{sec:dcrintro}. Without adjusting the model, the noise counts would cause the likelihood fit parameters to be biased. For example, the probability of a photon being detected before the first interaction in the ideal case is zero to a good approximation. A dark count that occurs before the first interaction would therefore force the best-fit interaction time to be equal to the time of this early dark count. Adding a small constant term to the likelihood model eliminates this problem with little effect on the resolution, assuming the dark count rate is low compared to the number of signal photons.  

Crosstalk photons have a more complicated effect on the likelihood fit, since they are highly correlated. Internal crosstalk acts essentially as a uniform scaling factor across the photon set, so it will follow the same time distribution and relative geometric distribution as the signal photons. It will therefore primarily affect the calculation of $\nu$. External crosstalk has the greatest potential to affect the fit results for the interaction location, time, and $\nu$. External crosstalk photons are likely to be detected in the opposite $8 \times 8$ SiPM array as the signal photon. The relative geometric distribution of external crosstalk will be nearly opposite that of the signal. The time distribution of external crosstalk will be delayed by the flight time of the optical crosstalk photons (about \SI{264}{ps} per \SI{50}{mm}) and will be additionally smeared by the SiPM time response function. These effects can, in principle, be accounted for in order to calculate a crosstalk profile for any proposed event. 

Adding terms for dark counts and optical crosstalk to the event likelihood gives the following:
\begin{equation}\label{eq:maxlik_ct}
\begin{split}
&  \quad \mathcal{L} = \frac{\nu^n}{n!}e^{-\nu}\prod_{i=1}^{n} \ \frac{1}{\nu}\ \Biggl[ \ \frac{R_{\alpha(i)}}{1-\mu_{\alpha(i)}}  \ + \\ 
&\ \sum_{j=1}^{N} \mathcal{N}_j \ \left( \mathcal{G}_{\alpha(i)}(\vec{x}_j) \ \mathcal{F}_{\alpha}(t_i-t_j-| \vec{x}_j-\vec{x}_{\alpha(i)} |/c_p) \ + \ \mathcal{C}_{\alpha(i)}\left( t_i; \ (\vec{x}_j,t_j) \right)  \right) \ \Biggr].
\end{split}
\end{equation}
The dark count time density, $R_{\alpha(i)}$, is equal to the constant DCR, accounting for the primary dark counts caused by thermal emission in pixel $\alpha(i)$ and contributions due to eXT from dark counts in all pixels (including second and higher order crosstalk from pixel $\alpha(i)$). The relationship between dark counts and eXT is complicated and the calculation is not straightforward. Fortunately, $R_{\alpha(i)}$ can be measured directly from dark count data. The scaling factor $\frac{1} {1-\mu_{\alpha(i)}}$ is the mean of the Borel distribution for iXT of pixel $\alpha(i)$, which has a Borel parameter of $\mu_{\alpha(i)}$. 

The term $\mathcal{C}_{\alpha(i)}\left( t_i; \ (\vec{x}_j,t_j) \right)$ is the non-normalized time-density function for crosstalk in pixel $\alpha(i)$, given a photon emitted isotropically from the $j^{th}$ interaction location $\vec{x}_j$ at time $t_j$. This function can be generated iteratively using the primary photon density, $P(\alpha(i),t_i;(\vec{x}_j,t_j) )= \mathcal{G}_{\alpha(i)}(\vec{x}_j)\cdot\mathcal{F}_{\alpha}(t_i-t_j-| \vec{x}_j-\vec{x}_{\alpha(i)} |/c_p)$ and a map of the pixel-to-pixel crosstalk probability,~$c_{\beta\alpha}$, which is defined as the probability that a crosstalk photon from a single photon detection in pixel~$\beta$ will be detected in pixel~$\alpha$. The first order approximation of the crosstalk density profile is given by a term for iXT, plus contributions for eXT from all other pixels:
\begin{equation}\label{eq:ctprof_1}
\mathcal{C}_{\alpha,1}\left( t; \ (\vec{x}_j,t_j) \right) = \mu_{\alpha}P'(\alpha,t-t_a;(\vec{x}_j,t_j))+\sum_{\beta \neq \alpha}c_{\beta\alpha}P'(\beta,t-t_a - | \vec{x}_{\beta}-\vec{x}_{\alpha} |/c_p;(\vec{x}_j,t_j)),
\end{equation}
where $t_a$ is the avalanche development time and the prime on $P'$ indicates that $P$ must be convolved with the SiPM time response. Higher order terms are then given by:
\begin{equation}\label{eq:ctprof_m}
\mathcal{C}_{\alpha,m>1}\left( t; \ (\vec{x}_j,t_j) \right) = \mu_{\alpha}\mathcal{C}_{\alpha,m-1}'\left( t-t_a; \ (\vec{x}_j,t_j) \right)+\sum_{\beta \neq \alpha}c_{\beta\alpha}\mathcal{C}_{\beta,m-1}'\left( t-t_a-| \vec{x}_{\beta}-\vec{x}_{\alpha} |/c_p; \ (\vec{x}_j,t_j) \right).
\end{equation}

The expected number of detected photons in \eqnref{eq:maxlik_ct} must now include contributions from dark counts and crosstalk, as well as primary scintillation photons. In general, $\nu$ will take the form:
\begin{equation}
\nu = \sum_{\alpha}R_\alpha t_w + \sum_{j}n_j \sum_{\alpha}\left[ \mathcal{G}_{\alpha}(\vec{x}_j) + \mu  \mathcal{G}_{\alpha}(\vec{x}_j)  + \sum_{\beta \neq \alpha}c_{\beta\alpha}\mathcal{G}_{\beta}(\vec{x}_j) + ...\right],
\end{equation} 
where $t_w$ is the width of the trigger window. In the special case where all pixels have the same iXT parameter ($\mu$) and eXT parameter ($c$), the expected number of detected photons will be given by:
\begin{equation}
\nu = \sum_{\alpha}R_\alpha t_w + \frac{1}{1-(\mu + c)}\sum_{j}n_j \Gamma(\vec{x}_j)
\end{equation}

The likelihood model described in \eqnref{eq:maxlik_ct} was tested using a simple Monte Carlo model of a 1D system of two pixels, where pixel~a is at $x=\SI{0}{cm}$, pixel~b is at $x=\SI{5}{cm}$, and the simulated source is at $x=\SI{2}{cm}$. The source generates a Poisson number of photons, with an average ($\nu$) of 100~photons per event. The photons are emitted according to a double exponential time profile, with the rise and fall times equal to datasheet values for EJ-204 (\SI{0.7}{ns} and \SI{1.8}{ns}, respectively)~\cite{ej204_datasheet}. We also simulate a \SI{200}{ps} timing resolution and define the speed of light in the medium to be \SI{30}{cm/ns}. We add an additional delay of \SI{0.5}{ns} to the external crosstalk photons.
We define a linear light collection ratio so that each photon has a probability of 3/5 of being detected in pixel~a and a probability of 2/5 of being detected in pixel~b. 
First, 5,000~events are simulated without dark counts or crosstalk and reconstructed by maximizing the likelihood described in \eqnref{eq:maxlik_ideal}. A second data-set is generated by adding SiPM noise to the same 5,000~events. The SiPM noise is simulated such that each pixel has an 
iXT parameter of 0.19, an eXT parameter of 0.25, and an expected 10 dark counts. This second data-set is reconstructed first using the ideal likelihood shown in \eqnref{eq:maxlik_ideal} and then using the likelihood accounting for SiPM noise shown in \eqnref{eq:maxlik_ct}. 

\begin{figure}[htbp]
\centering 
\includegraphics[width=.9\textwidth]{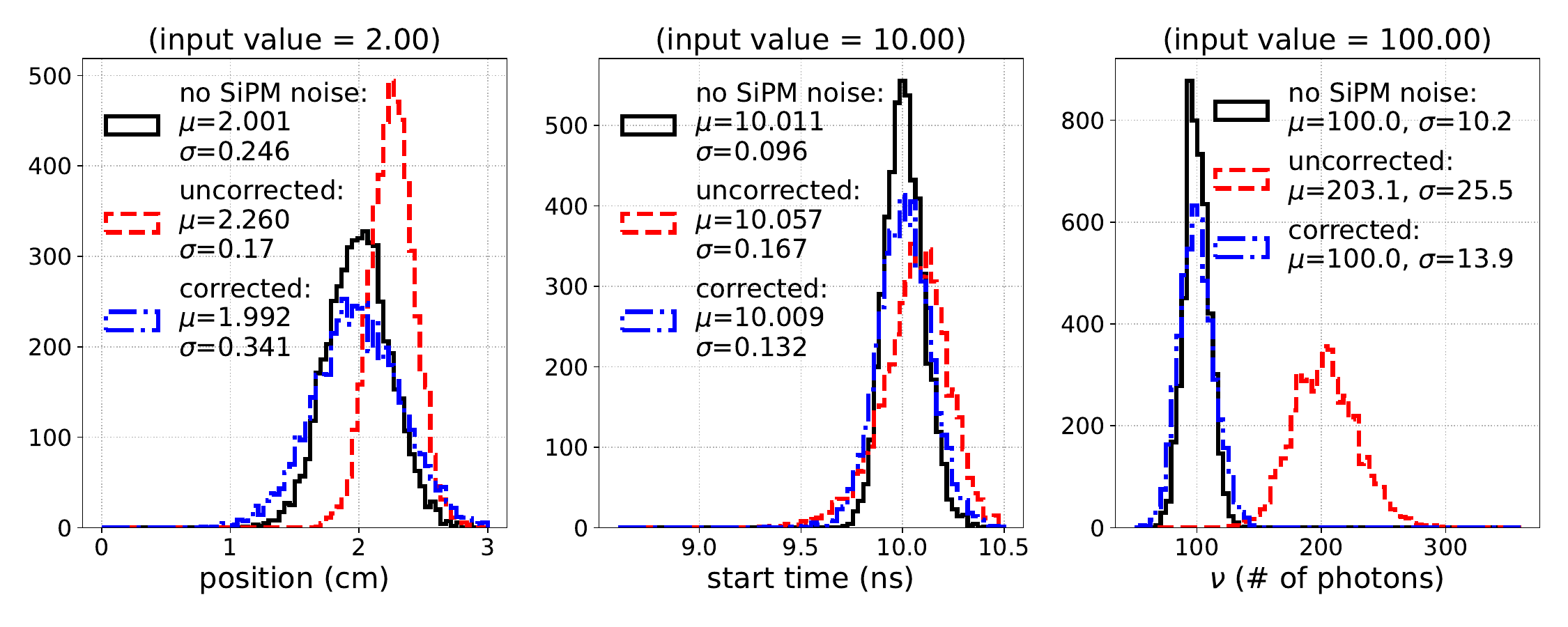}
\caption{\label{fig:2pixres}Histograms of maximum-likelihood parameters for simulated event reconstruction in a 1D two-pixel system as described in the text. The individual panels show the reconstructed values of the source location (left), the pulse start time (middle), and the expected number of photons (right). The three histograms in each panel show the cases of 5,000 simulations without SiPM-noise reconstructed using the ideal likelihood model (black), 5,000 simulations with SiPM-noise reconstructed using the ideal likelihood model (red), and 5,000 simulations with SiPM-noise reconstructed using a likelihood model that accounts for SiPM-noise (blue). }
\end{figure}

\begin{table}[!htbp]
\centering
\caption{\label{tab:mcres}Results table for the reconstruction of the two-pixel Monte Carlo model described in the text. We list means and widths of the associated distributions shown in \fig{fig:2pixres}, along with their uncertainties.}
\begin{tabular}{|| c || c || c | c || c | c || c | c ||}
\hline
 \multicolumn{2}{| c ||}{} & \multicolumn{2}{c ||}{\textbf{Ideal MC}} & \multicolumn{2}{c ||}{\textbf{SiPM MC}} 
& \multicolumn{2}{c ||}{\textbf{SiPM MC}}\\
 \multicolumn{2}{| c ||}{} & \multicolumn{2}{c ||}{\textbf{Ideal Recon.}} & \multicolumn{2}{c ||}{\textbf{Ideal Recon.}} 
& \multicolumn{2}{c ||}{\textbf{SiPM Recon.}}\\
\hline
\textbf{Parameter} &  \textbf{Input}  &  \textbf{mean} & \textbf{width} &  \textbf{mean} & \textbf{width}    &   \textbf{mean} & \textbf{width}   \\
\hline
Source Pos.    &  2  & 2.001 &  0.246 &  2.260 &  0.170 &  1.992 &  0.341 \\
(cm)                &  {} & $\pm$0.003 &  $\pm$0.002 &  $\pm$0.002 &  $\pm$0.002 &  $\pm$0.005 &  $\pm$0.003 \\
\hline
Pulse Time   &  10  & 10.011 &  0.096 &  10.057 &  0.167 &  10.009 &  0.132 \\
(ns)                &  {} & $\pm$0.001 &  $\pm$0.001 &  $\pm$0.002 &  $\pm$0.002 &  $\pm$0.002 &  $\pm$0.001 \\
\hline
$\nu$  &  100  & 100.0 &  10.2 &  203.1 &  25.5 &  100.0 &  13.9 \\
(photons)    &  {} & $\pm$0.1 &  $\pm$0.1 &  $\pm$0.4 &  $\pm$0.3 &  $\pm$0.2 &  $\pm$0.1 \\
\hline

\end{tabular}
\end{table}

Histograms of the reconstructed parameters are shown in \fig{fig:2pixres}, and the results are listed in \tab{tab:mcres}. For the ideal data-set, the reconstructed values of source position and $\nu$ are consistent with the input values. There is an \SI{11(1)}{ps} bias in the pulse start measurement. The origin of this bias is unknown, but it is small compared to the simulated timing resolution of \SI{200}{ps}. In the SiPM-noise data-set, the parameters are all reconstructed with significant bias when using the ideal likelihood. When the SiPM-noise data-set is reconstructed instead using the likelihood proposed in \eqnref{eq:maxlik_ct}, the added bias is removed, and the reconstructed parameters are consistent with those from the ideal data-set. 

The likelihood model proposed in \eqnref{eq:maxlik_ct} allows bias due to SiPM noise to be removed from the reconstructed parameters. However, the resolution of the unbiased parameters will be degraded versus the ideal case. The standard deviations of the distributions shown in \fig{fig:2pixres} for the parameters reconstructed using the corrected likelihood model are all increased by about 38\% over the ideal data-set.

\section{Single Photon Processing}\label{sec:sptr}
Three digital signal processing (DSP) steps are applied to the raw waveforms to enable single-photon analysis. The waveforms are first resampled to mask various pathologies of the SCEMAs and to provide a trace with a uniform sampling period. A digital band-pass filter is applied to the resampled waveforms to better isolate individual photon peaks. Finally, a peak-finding algorithm is applied that yields the timing and amplitudes of all the photon peaks in the waveform.

\begin{figure}
\centering
\includegraphics[width=.8\textwidth]{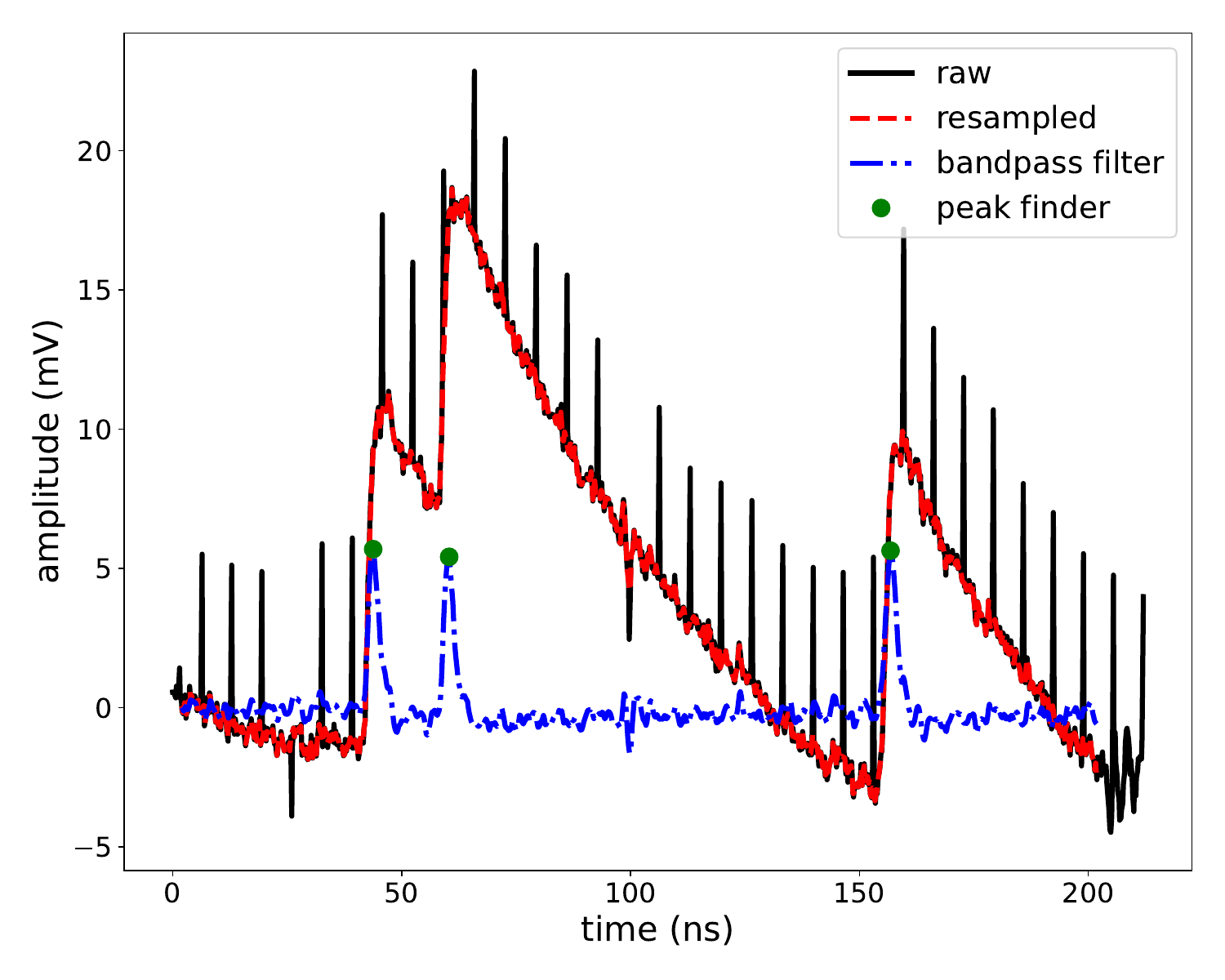}
\caption{\label{fig:traceprep} Example of the pulse processing performed on the single photon pulses. The black curve shows the digitized output of a SCEMA prior to any pulse processing. The red dashed line shows the same trace after the 32-bin spikes are removed and the timing has been resampled to a uniform rate of \SI{5}{GHz}. The blue dot-dashed curve shows the output of a band pass filter applied to the resampled trace. The green markers show the amplitude and locations of the peak-finder output.
}
\end{figure}

There are several pathologies observed in the SCEMA output waveforms that must be removed before performing any further DSP steps. The ``raw'' trace in \fig{fig:traceprep} shows an example of the unmasked output of the SCEMAs. The most obvious pathology is the large periodic spikes that occur throughout the trace. These spikes likely stem from a hardware issue on the DRS4s as they are always associated with specific DRS4 capacitors (or time bins). The indices of these capacitors are always equal to 30 or 31, modulo 32. Another pathology in the SCEMA waveforms can be seen in the final \SI{20}{ns} of \fig{fig:traceprep}. The baseline near the end of the trigger window begins to trend upward, and the noise increases dramatically. The time bins that exhibit these pathologies are masked from the final waveform in software. The first \SI{2}{ns} of each trace is also masked to prevent a similar baseline issue that can occur at the beginning of the trigger window. The total size of the SCEMA trigger window is about \SI{220}{ns}, so after our cuts we are left with a \SI{198}{ns} region which we resample using linear interpolation to a uniform \SI{0.2}{ns} sample period. 

As shown in \fig{fig:traceprep}, single photon waveforms have tails that span tens of nanoseconds that can affect measurement of the amplitude and timing of subsequent pulses. The band-pass Butterworth filter from the {\tt scipy.signal} package~\cite{scipy} is applied to the raw trace to help resolve individual photon peaks. Optimization of the window was performed for a SiPM biased at \SI{57}{V}, with an RF amplifier bias of \SI{9}{V}. The cutoff frequencies were selected by hand to minimize the width of a single photon pulse, while maintaining a high signal-to-noise ratio. The frequency window \SIrange{100}{600}{MHz} is selected, which yields a single-photon peak with full-width-half-max of about \SI{2.6}{ns} and a signal-to-noise ratio of about 21.

\begin{figure}
\centering 
\includegraphics[width=.8\textwidth]{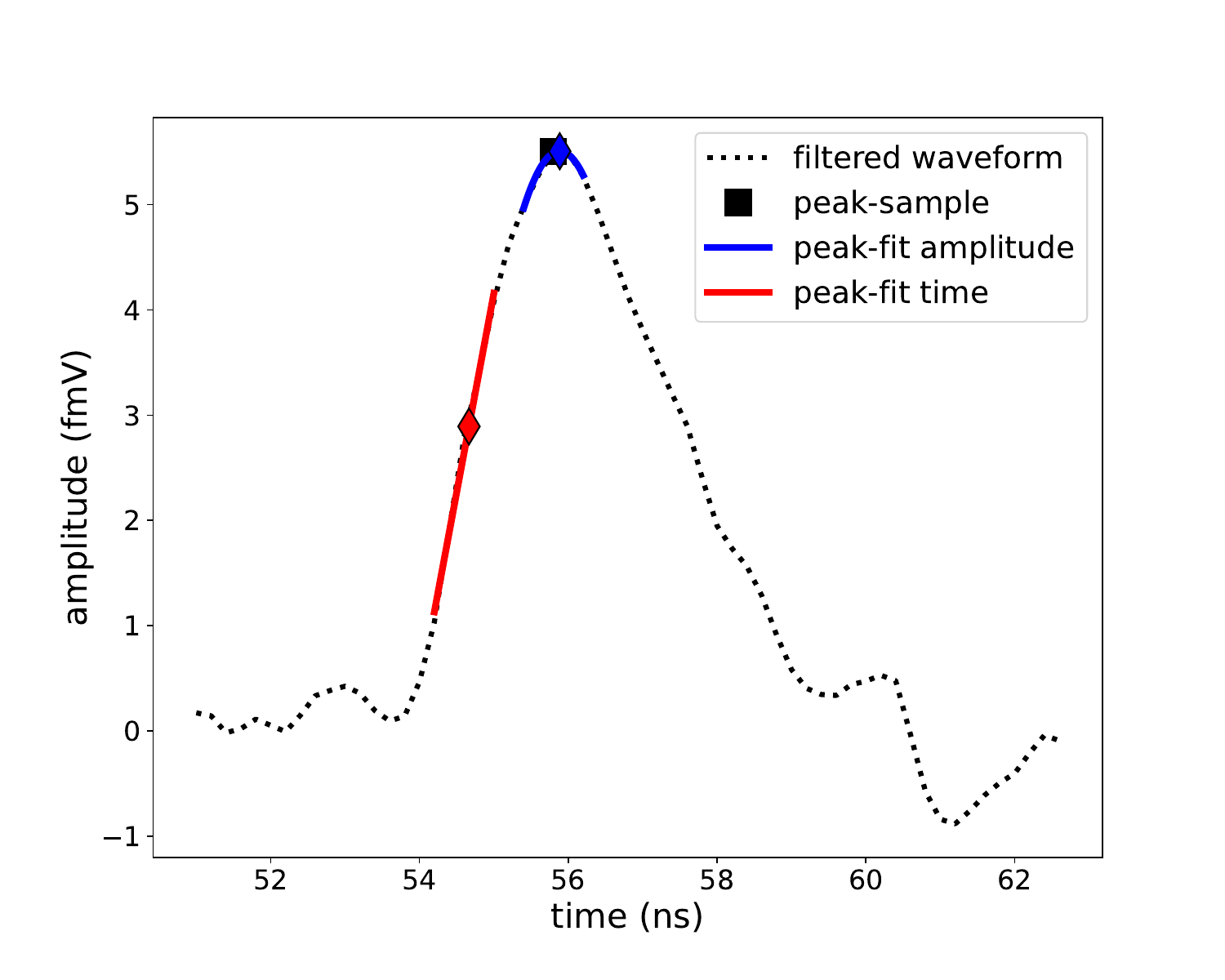}
\caption{\label{fig:peakfit} Example of the peak analysis performed on the SiPM pulses. The black dotted line shows the bandpass-filtered trace. The red line is fit to the rising edge of the pulse, and the peak-fit time is taken to be the time at which this line crosses \SI{50}{\percent} of the peak-fit amplitude. The blue line show a quadratic fit to the top of the pulse. The peak-fit amplitude is taken to be the maximum value of the quadratic function. The black square shows the time and amplitude of the maximum sample in the pulse (i.e. the peak-sample time and amplitude).
}
\end{figure}

The peak-finding algorithm is applied to the resampled and band-pass filtered waveform. First, a selection cut is made on waveform samples that are greater than a threshold of \SI{1.5}{\au} (arbitrary units corresponding to filtered millivolts), and the selected samples are iterated over. 
This threshold is roughly five times the standard deviation of the baseline and one third of the single photon amplitude. The time and amplitude of the maximum sample of each rising edge are defined as the peak-sample values. All peaks are required to be separated by at least two samples (\SI{0.4}{ns}). In addition to peak-sample information, polynomial fits are applied to calculate the peak time and amplitude as is shown in \fig{fig:peakfit}. A quadratic equation is fit to the five nearest samples to the peak-sample value, and the maximum value of the best fit quadratic is taken as the peak-fit amplitude. A linear fit is then applied to four samples along the rising edge of the peak. The time at which this line crosses \SI{50}{\percent} of the peak-fit amplitude is taken to be the peak-fit time. 

The DSP chain described above is validated using data taken with a Photek~LPG-405 pulsed laser~\cite{lpg405} incident on a SiPM biased at \SI{57}{V} with an RF amplifier bias of \SI{9}{V}. The laser is attenuated using neutral density filters such that fewer than one photon per pulse is expected to be detected while maintaining a high rate compared to dark counts. The laser pulse has a quoted width of \SI{75}{ps~FWHM}. For this test, the SiPM timing is aligned to the sync pulse from the laser.
In addition to the peak-sample values and peak-fit values, peak-template values are extracted using a template fitting method. 
The template used in this method is taken from the central band of many single-photon waveforms. 
The waveforms are first aligned such that the peak-sample times for each waveform is equal to \SI{0}{ns}.
The samples from all the waveforms are combined to give the two-dimensional (time-vs-amplitude) single-photon waveform distribution.
A preliminary single-photon template is measured by binning this distribution in time and calculating the central value in each time bin. 
The preliminary template is fit to each of the single-photon waveforms, floating the time-offset and amplitude of the template. 
The time axis for each waveform is re-aligned such that the best-fit times are equal to \SI{0}{ns}.
The template measurement is repeated on the re-aligned set of waveforms to give the final single-photon template.
The final template is fit to all of the peaks in a data-set by floating the amplitude and time offset, yielding the template-fit amplitudes and times. 
This method utilizes all of the samples in the waveform to measure amplitude and timing and so will minimize statistical uncertainty in the amplitude and time values, assuming a consistent waveform shape.

\begin{figure}
\centering 
\includegraphics[width=.8\textwidth]{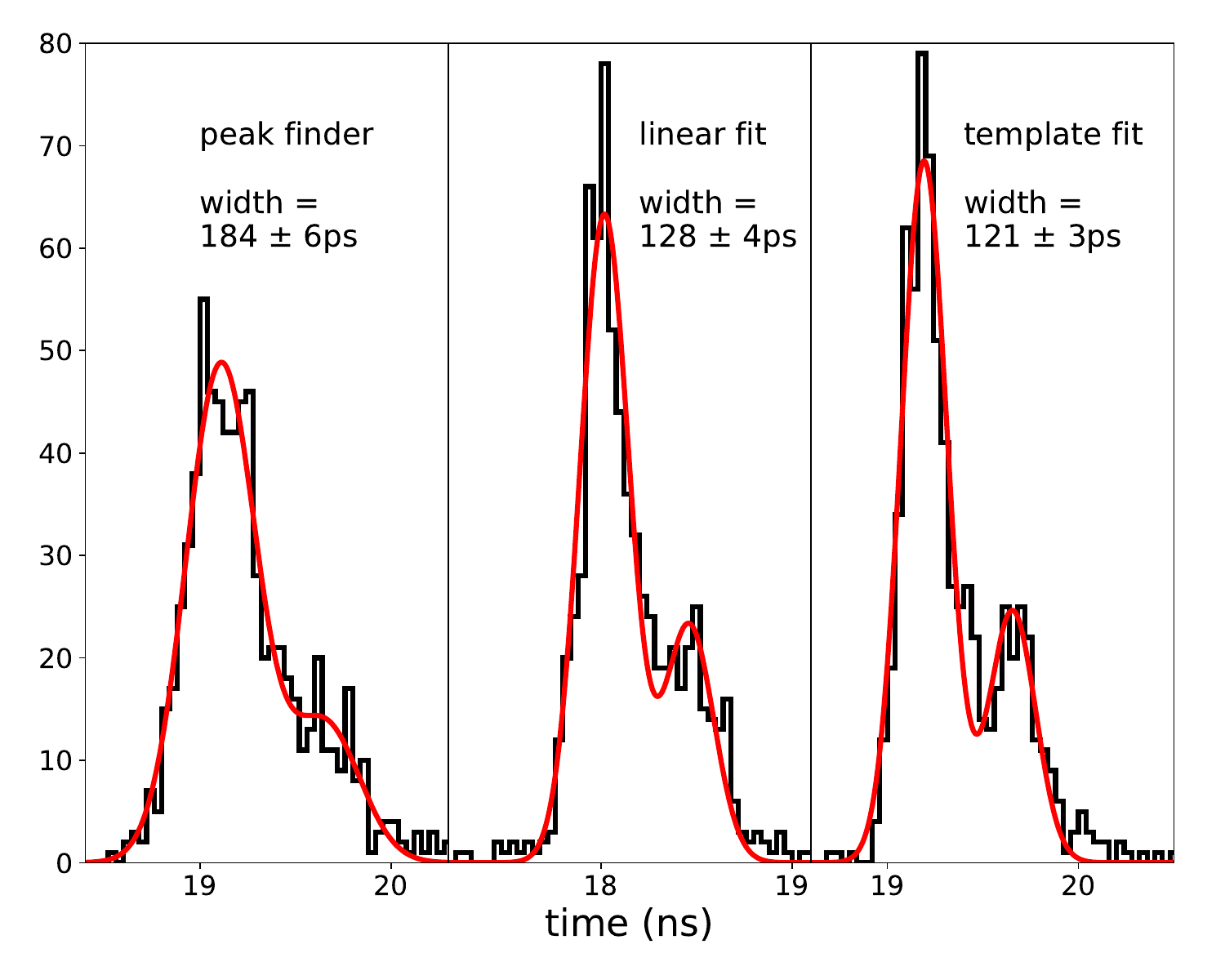}
\caption{\label{fig:sptr} Histograms (black lines) of the measured arrival times of photons using three different analysis methods. The photons are from a Photek LPG-405 pulsed laser, and the time is measured relative to the laser sync pulse. The fitting model (red lines) used is the sum of two Gaussian functions. The left hand plot shows arrival times taken simply from the time of the max sample in the peak (peak-sample times). The middle plot shows arrival times measured from linear fit to the rising edge of the peak (peak-fit). The right hand plot shows arrival times measured using a template-fitting method described in the text. }
\end{figure}

The laser used is known to produce two prominent peaks in the arrival-time distribution separated by several hundred picoseconds, so the sum of two Gaussian functions is fit to the measured spectra. Both Gaussian functions in each fit are required to have the same width. 
The best fit Gaussian standard deviations are shown in \fig{fig:sptr}. The peak-sample timing yields the broadest spectrum at \SI[separate-uncertainty = true]{184(6)}{ps}. The polynomial and template fits each yield spectra with comparable widths at \SI[separate-uncertainty = true]{128(4)}{ps} and \SI[separate-uncertainty = true]{121(3)}{ps}, respectively. 
The template-fitting method takes about an order of magnitude more computing time to complete than the polynomial peak-fitting method, with minimal improvement to the timing resolution.

\begin{figure}
\centering 
\includegraphics[width=.8\textwidth]{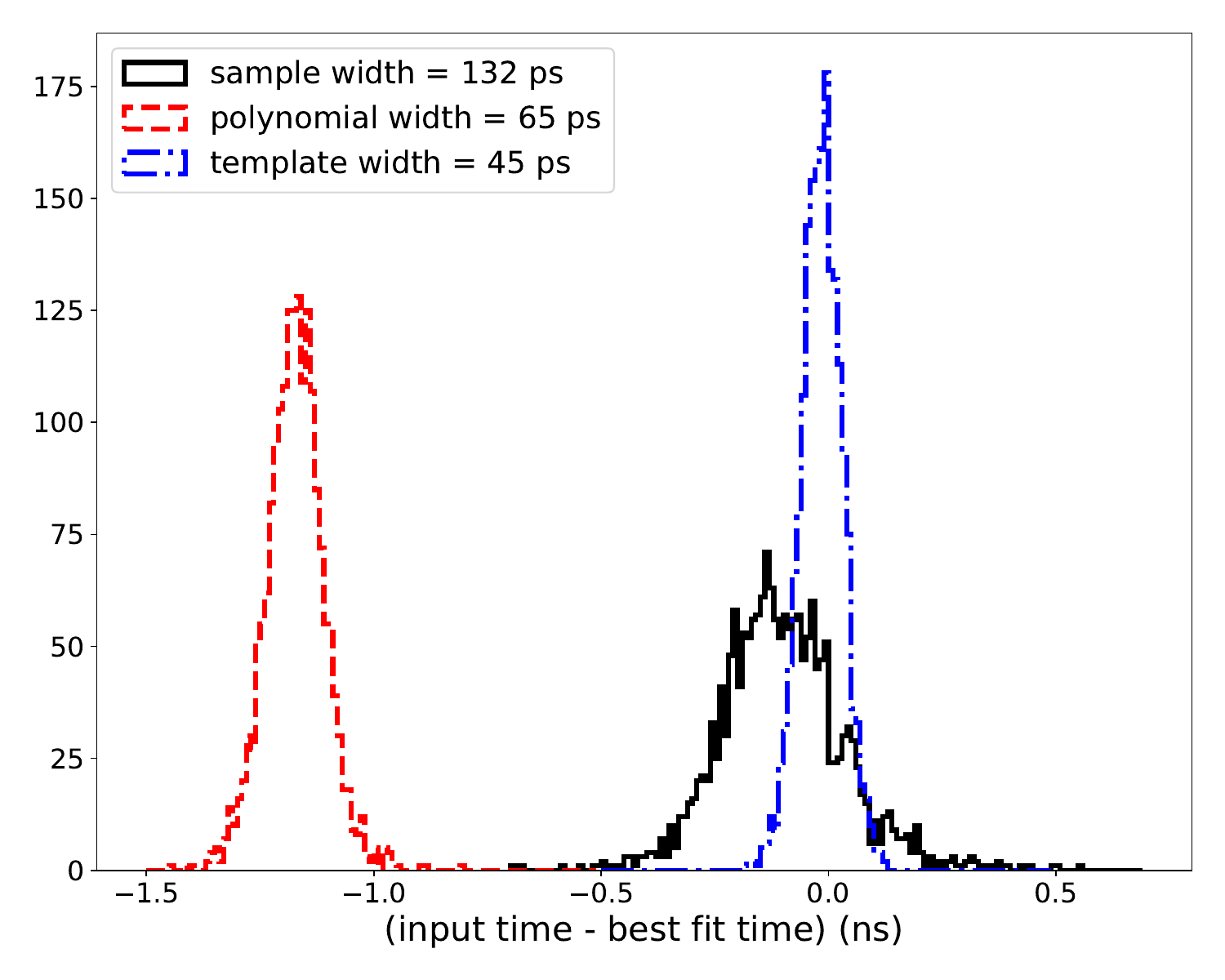}
\caption{\label{fig:MCsptr} Measured time vs input time for simulated pulses. The pulse shape used in the simulation is measured from single-photon pulses in the data. The arrival times are measured using the maximum sample time (black), a linear fit to the rising edge (red), and a template fitting method (blue).}
\end{figure}

The trigger time is measured using a linear fit to the rising edge of the digitized trigger pulse, and its contribution to the measured 1-$\sigma$ photon timing resolution is expected to be \SI[separate-uncertainty = true]{25(11)}{ps}. The expected contribution to timing resolution due to the DSP and analysis is shown in \fig{fig:MCsptr}. These contributions are estimated using a Monte-Carlo method in which simulated single photons are injected into SCEMA traces that have no peaks. The DSP steps described above are applied to the simulated photons, and the difference between the known photon time and the measured photon time is shown in \fig{fig:MCsptr}. The peak-sample analysis is expected to contribute about \SI{132}{ps}, the polynomial fitting method will contribute about  \SI{65}{ps}, and the template fitting method will contribute about \SI{45}{ps}. The remaining width of the timing spectra is due to the inherent single-photon timing resolution (SPTR) of the SiPMs and readout circuitry. If we assume that these contributions are uncorrelated or normally distributed, we calculate an inferred SPTR of about \SI{105}{ns} using the template-fitting analysis, which is in line with what was measured in \cite{cates2022}. 

\section{Dark Count Analysis}\label{sec:darkcounts}

Event reconstruction through maximization of the likelihood function in \eqnref{eq:maxlik_ct} requires knowledge of several detector-dependent parameters. Ideally, these parameters would all be measured directly using calibration data.
The parameters are listed in Appendix~\ref{app:parms}, along with their descriptions and how they will be characterized. The pulse height spectrum from a set of dark count data taken over a three-day period is used to calculate the single photon pulse-height, as well as the Borel parameter for iXT. This data-set was triggered using random triggers from a DG535 so that the timing of the eight SCEMAs could be aligned, thus allowing time-correlation analysis of dark counts across channels. The time-correlation analysis allows for in-situ validation of the timing resolution, as well as direct measurement of the pixel-to-pixel eXT probability.

\begin{figure}
\centering 
\includegraphics[width=.99\textwidth]{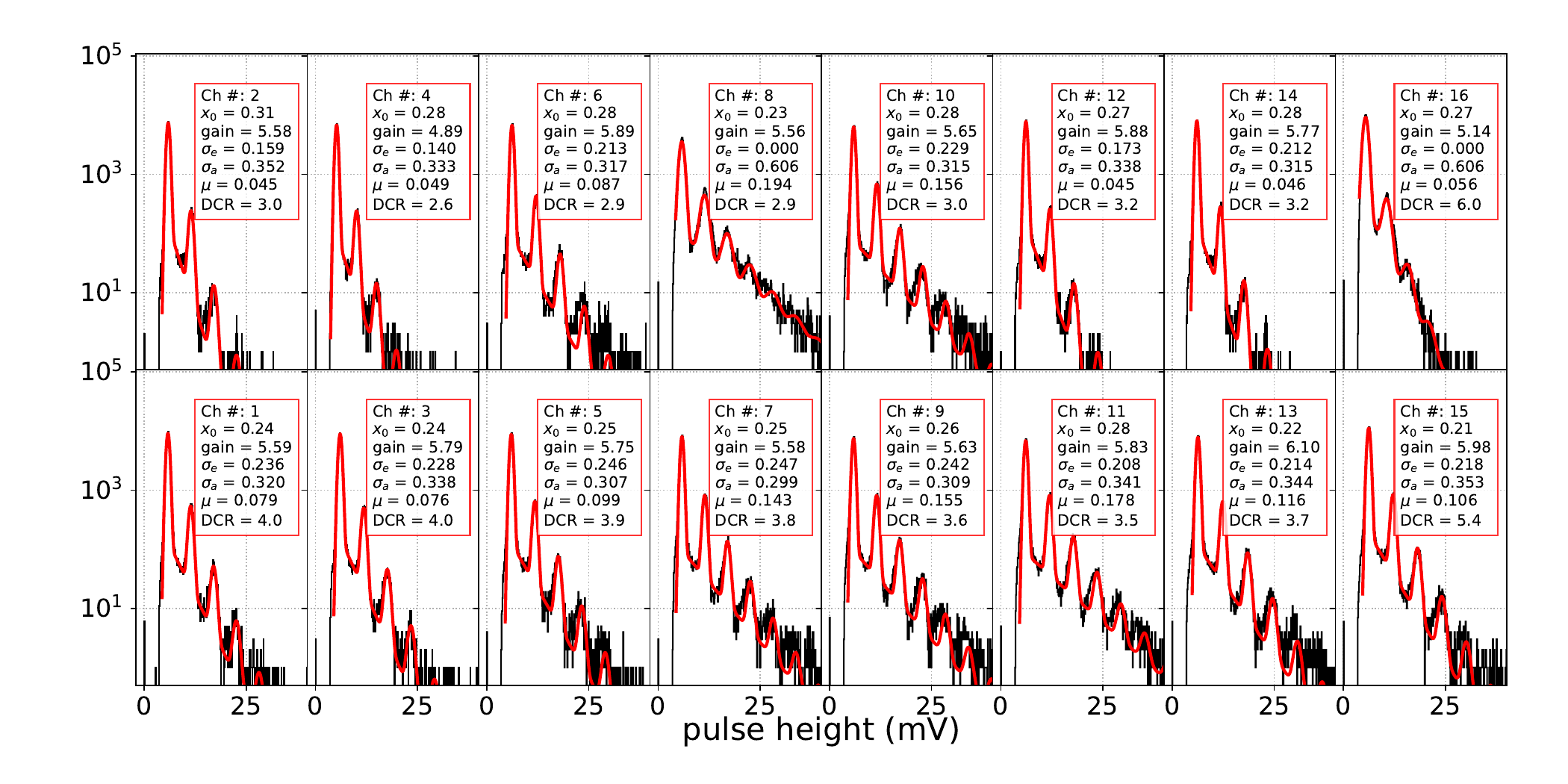}
\caption{\label{fig:dcrfit0} Fits to the pulse-height spectra measured from the dark count data-set for all channels on SCEMA board 0. The fitting model is described in \secref{sec:dcrintro}. The gain has units of filtered-mV per photon, the crosstalk parameter is unitless, and the dark count rate has units of MHz.  }
\end{figure}

The single photon processing as described in \secref{sec:sptr} is applied to the dark count data-set described above. The resulting list of peak amplitudes gives the single-photon pulse-height spectrum (PHS). The PHS for each pixel is fit according to \eqnref{eq:ser_gausfit} to measure the average single-electron pulse height ($x_g$), the width due to electronic noise ($\sigma_e$), and the width due to avalanche fluctuations ($\sigma_a$). 
As an example, the fits for each SiPM channel on SCEMA board~0 are shown in \fig{fig:dcrfit0}. Fits for the remaining boards are shown in Appendix~\ref{app:dcrfits}.

The total DCR for a given pixel, accounting for eXT, is equal to the total number of peaks over a threshold of \SI{3.5}{mV}, divided by the live time. In this case, each of the 130,000 trigger windows has a live time of \SI{198}{ns}, giving a total live time of \SI{25.7}{ms}. 

The Borel parameter for iXT is also measured from the PHS fits, but is not used here. Instead, the iXT probability is defined as the probability that a single photon pulse will have an avalanche multiplicity of at least~2 ($P_{iXT} \equiv N_{M\geq 2}/N_{tot}$). The number of photon pulses with a multiplicity of at least~2 ($N_{M\geq 2}$) is calculated by counting the number of peaks with amplitudes greater than $x_0 +1.5\cdot x_g$. This value is then divided by the total number of peaks ($N_{tot}$) in the channel to give~$P_{iXT}$.

\begin{figure}
\centering 
\includegraphics[width=.99\textwidth]{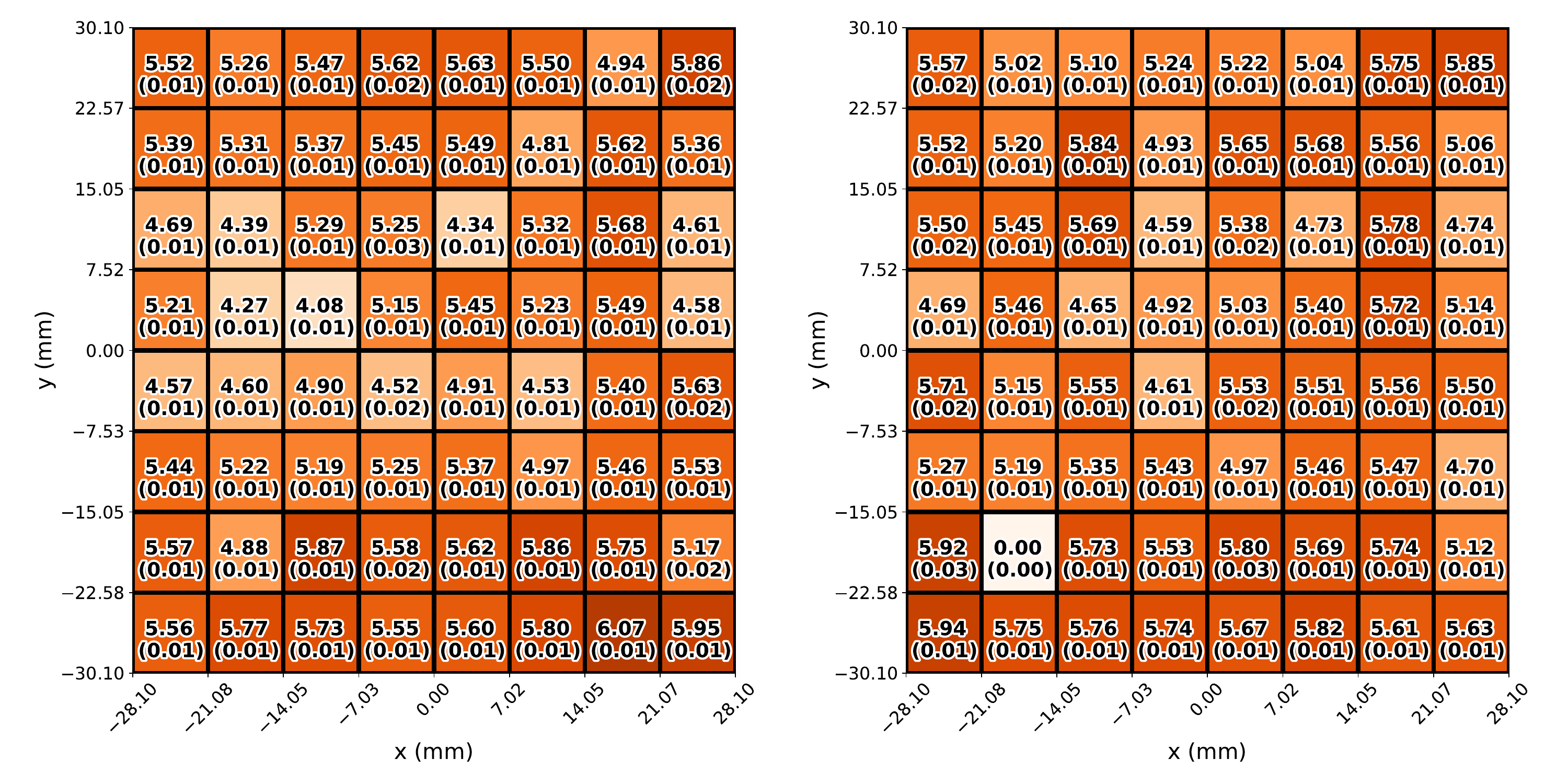}
\caption{\label{fig:gainmap} Average single-electron pulse height in millivolts for each SiPM channel. The fitting uncertainty is shown in parentheses. Channel 86 is non-functional as indicated by a 0 result.  }
\end{figure}

\begin{figure}
\centering 
\includegraphics[width=.99\textwidth]{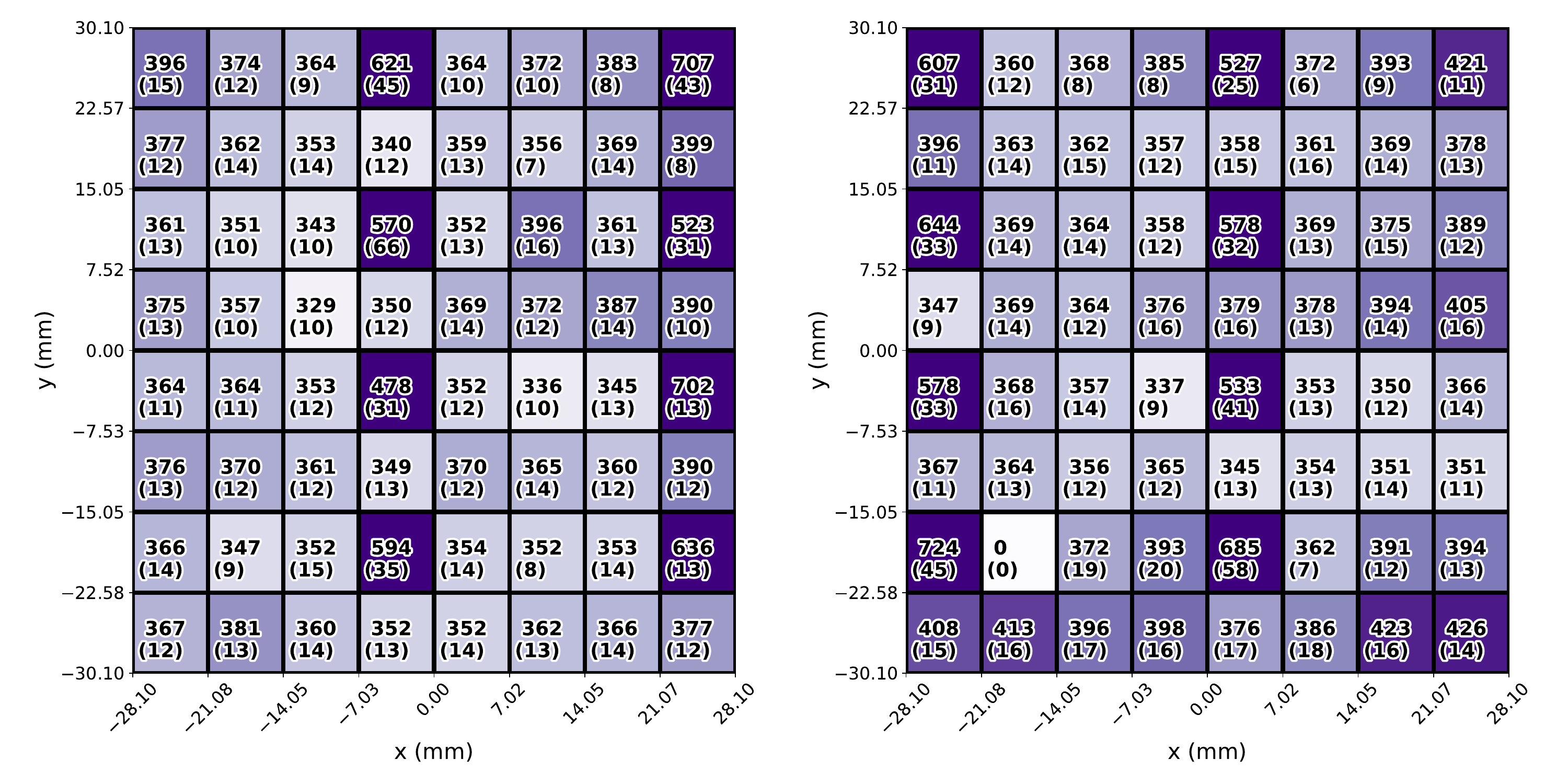}
\caption{\label{fig:spsigmap} Average single-electron width in microvolts for each SiPM channel. The fitting uncertainty is shown in parentheses. Channel 86 is non-functional as indicated by a 0 result.  }
\end{figure}

\begin{figure}
\centering 
\includegraphics[width=.99\textwidth]{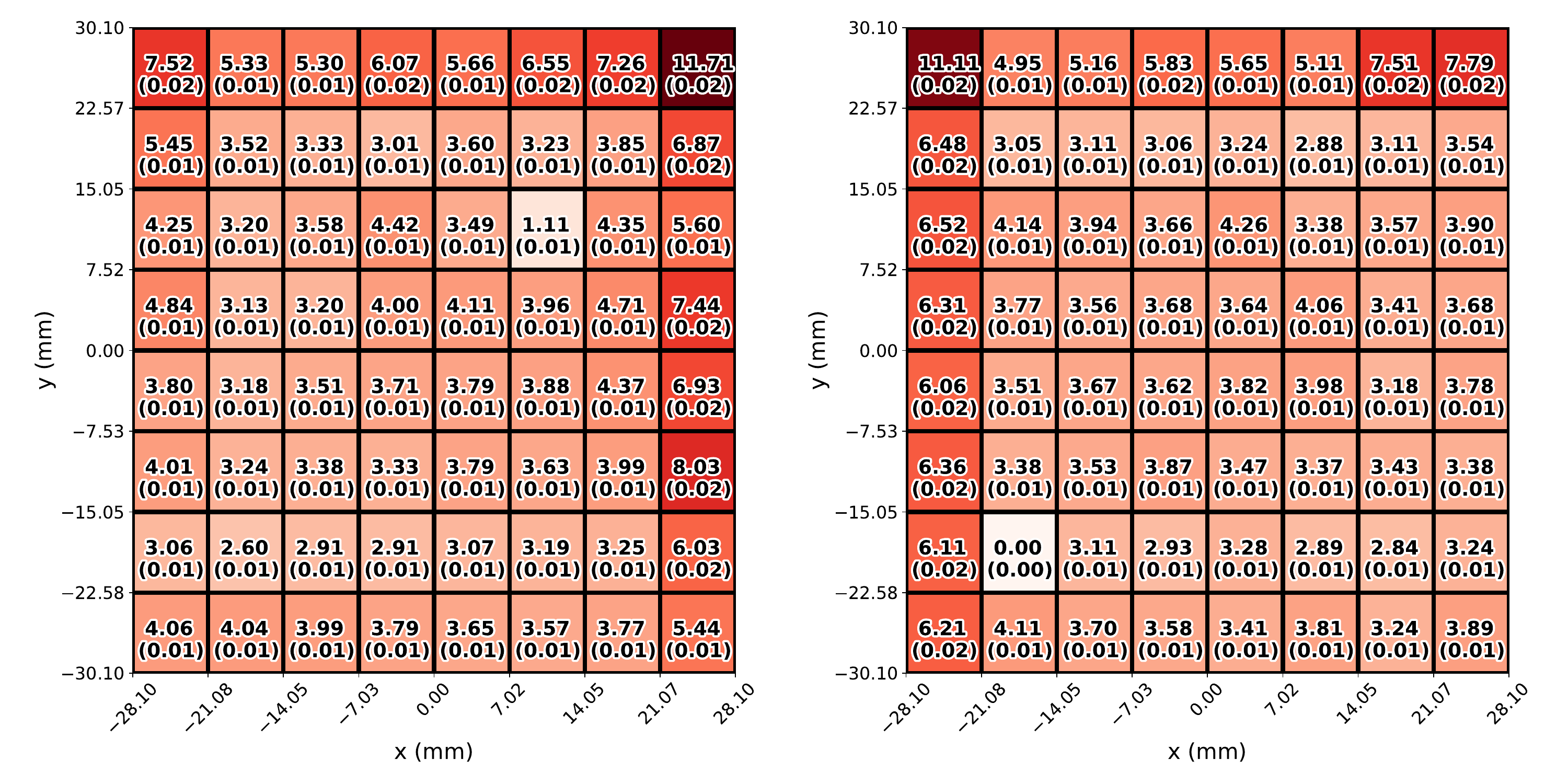}
\caption{\label{fig:DCRmap} The measured dark count rate in MHz for each SiPM channel. The uncertainty due to Poisson counting is shown in parentheses.  Channel 86 is non-functional as indicated by a 0 result. }
\end{figure}

\begin{figure}
\centering 
\includegraphics[width=.99\textwidth]{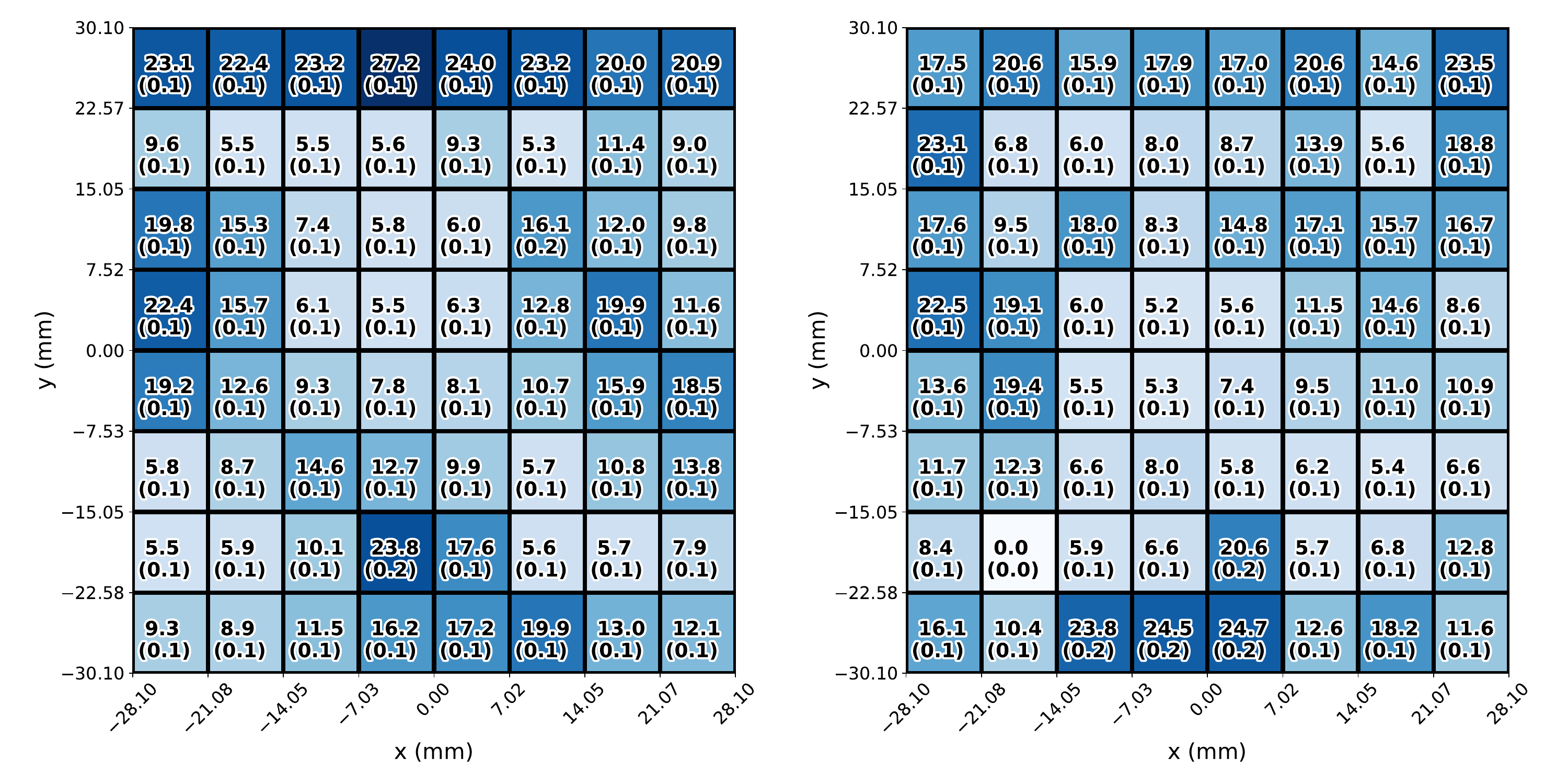}
\caption{\label{fig:XTmap} The measured iXT, in units of percent, for each SiPM channel. The uncertainty due to Poisson counting is shown in parentheses. Channel 86 is non-functional as indicated by a 0 result.}
\end{figure}

The results from the PHS fits are arranged into heat-maps and shown in Figures~\ref{fig:gainmap},~\ref{fig:spsigmap},~\ref{fig:DCRmap}, and~\ref{fig:XTmap}.
The SiPM channel assignments are the same as those shown in \fig{fig:channellayout}. There are 2 out of the 128 channels that are found to be either non-functional or semi-functional. Channel 48 functions normally for about half of the files but is non-functional in the other half, and channel 86 is entirely non-functional.

The average single-electron sizes ($x_g$) for each pixel are shown in \fig{fig:gainmap}, and the single electron widths ($\sqrt{\sigma_a^2+\sigma_e^2}$) are shown in \fig{fig:spsigmap}. There is a significant amount of variation in the gains, which is likely dominated by the nonuniform breakdown voltages across the SiPM channels. The average gain is \SI{5.3}{mV}, with a standard deviation between the channels of \SI{0.4}{mV}. There is no clear geometry dependence to the gain variation that would indicate a problem with the front-end or SCEMA circuits. 

The best-fit single electron widths show a clear geometry dependence. SiPMs read out by channels~8 and~16 on each SCEMA measure single-electron widths up to two times larger than the other channels. Typical channels have an average single-electron width of \SI{379}{\mu V}, with a standard deviation of \SI{5}{\percent} between channels. The high-width channels have an average single-electron width of \SI{607}{\mu V}, with a standard deviation of \SI{12}{\percent}. This effect also causes our fitting model to fail by forcing $\sigma_e =$~\SI{0}{mV} in most of the affected channels. We are unsure what is causing these channels to behave differently. There are no obvious irregularities in the associated traces on the front-end boards or the SCEMA boards.

\begin{figure}
\centering 
\includegraphics[width=.8\textwidth]{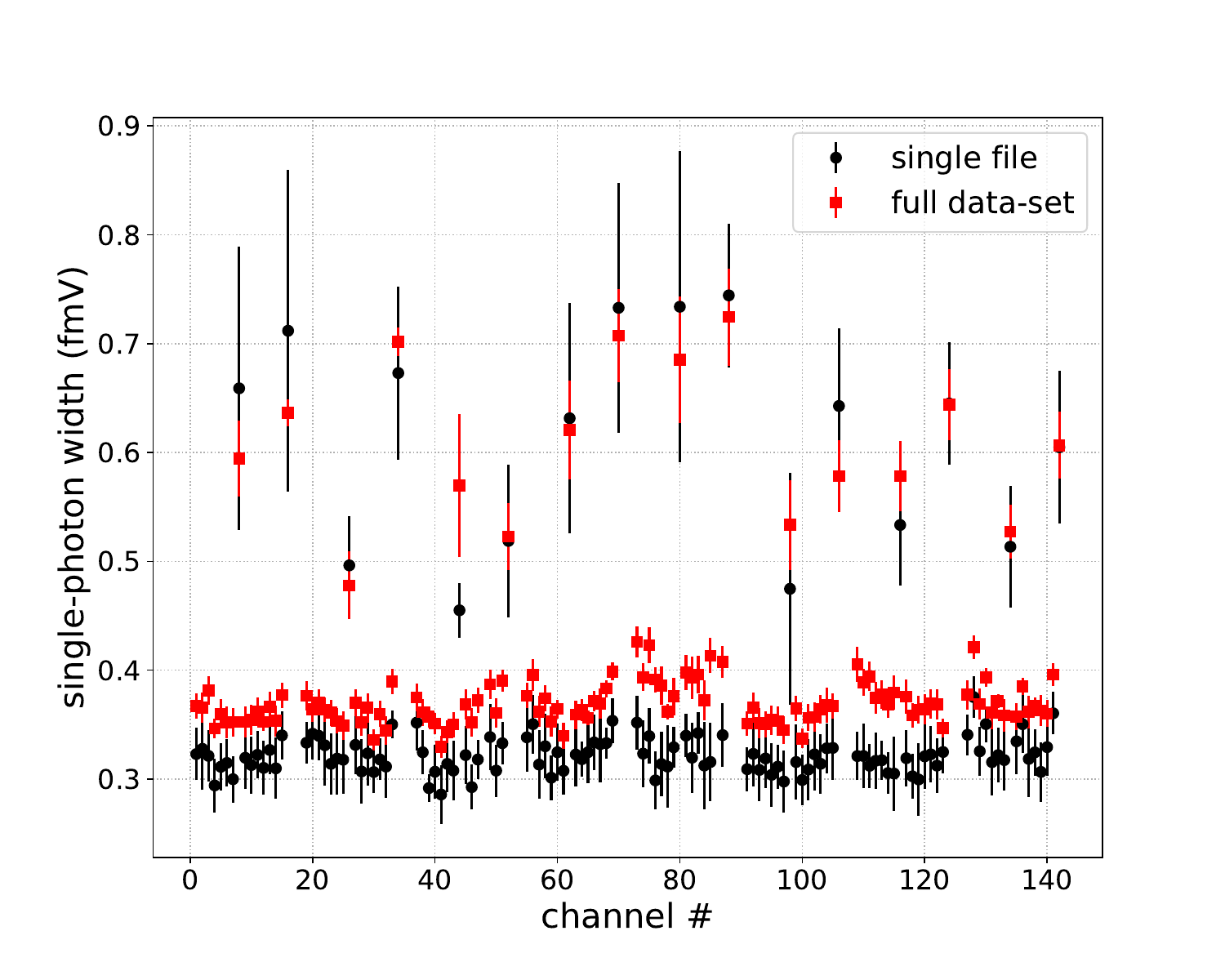}
\caption{\label{fig:SPwidth} Measured width of the single photon peaks for all SiPM channels. The black 
circles show the width measured from a single file, containing data from a roughly 30 minute run. The 
red squares show measurements using the full data-set, which contains about 13 hours of data taken 
over a 3 day period. The error bars indicate the fitting uncertainty for each value. }
\end{figure}

The single-electron widths shown in \fig{fig:spsigmap} are measured using the full data-set, which 
contains about 13 hours of data taken over a 3 day period. In \fig{fig:SPwidth}, these are 
compared to widths measured using data from a single file, which contains about 30 minutes of data. 
Comparing the widths of the full data-set peaks to the single-file peaks gives a measure of how much gain 
drift affects the final amplitude resolution. Excluding the channels with elevated widths, the average peak-width for the full data-set is
\SI{0.369}{\au} with a standard deviation of \SI{0.019}{\au}, and the average peak-with for the single-file selection is \SI{0.321}{\au} with a standard deviation of \SI{0.016}{\au}. 
The average measured difference of \SI{0.048}{\au} is larger than the average combined fitting error of \SI{0.029}{\au}, and indicates about a \SI{15}{\percent} increase over the single-file subset. 

The measured DCR in MHz and Poisson counting uncertainty for each SiPM is shown in 
\fig{fig:DCRmap}. The threshold of \SI{3.5}{mV} is \SI{4.7}{\sigma} below the single-photon peak for 
a typical SiPM, so there will be negligible loss in dark counts due to thresholding. SiPMs on SCEMA 
channels~8 and~16 have slightly higher leakage due to their wider single electron peaks. The maximum 
loss in dark counts due to thresholding is expected to be \SI{4}{\percent}. The DCR per channel ranges from 
\SI{2.6}{MHz} to \SI{11.7}{MHz}, with an average value of \SI{4.3}{MHz} and standard deviation of 
\SI{1.5}{MHz}. The measured iXT parameters, along with uncertainty due to Poisson counting, are shown in \fig{fig:XTmap}.

The SiPMs located at the $+y$ edge of the detector have elevated DCR and iXT. This is likely, in part, due to the SiPM array overhanging the scintillator, leaving some SiPM active area exposed to the air. This overhang would cause elevated DCR in two ways. First, the overhanging SiPM surface would be exposed to any light leakage into the dark box. Second, the lack of index-matching on the overhanging surface could increase reflections, redirecting some eXT photons back into the primary pixel~\cite{Masuda2021}. Some of these photons would cause prompt signals, increasing the measured iXT, and some would cause delayed signals, increasing the measured DCR. 
There is not a similar effect seen in the $-y$ edge although it has a similar overhang as the $+y$ edge. This difference could point to non-uniformity in how the optical interface pad is seated to the SiPM array, although we were not able to confirm this after visual inspection. This effect was not anticipated prior to assembly, and we do not have a complete understanding of its origin. This unpredictability illustrates the importance of the in-situ characterization described in this section. 

We observe increased DCR in the $+x$ edge of the left array and the $-x$ edge of the right array in \fig{fig:DCRmap}. We believe that this effect is due to peculiarities in the circuit design because it was observed in testing of the SiPM arrays prior to assembly. These pixels are not expected to have significant overhang, and the iXT in these SiPMs is not elevated, indicating that the elevated DCR in these channels may be caused by a different effect than the SiPMs at the $+y$ edge.

\section{Time-Correlation Analysis}\label{sec:timecorr}
Thermal emission in SiPMs is typically understood to be an exponential process, which is uncorrelated in time~\cite{acerbi2019}. Any measured time coincidence in the dark count data set described in \secref{sec:darkcounts} is therefore taken to be due to systematic correlation effects. The dominant source of time correlation between SiPMs is expected to be eXT. Other expected sources are electrical crosstalk and external coincidence such as light leakage and background scintillation. Electrical crosstalk in the SiPM arrays has been measured to be \SI{<5}{\percent} between adjacent channels and \SI{<0.1}{\percent} between all other channel pairs. Single-photon signals are therefore not expected to produce measurable electrical crosstalk. The SiPMs are well isolated from external light sources, as described in \secref{sec:detector}, so light leakage should be rare. Any external photons that do reach the SiPMs will not be time-correlated on the nanosecond scale and so can be treated the same as thermal emission for the purpose of this analysis. We do not expect any large background scintillation signals within the \SI{25.7}{ms} live-time of our dark count data set, and we do not observe signs of significant scintillation in the measured pulse-area histogram. To be safe, we do not include in the time-correlation analysis any events with a pulse-area summed across all 128 channels greater than 200 times the single-photon area.

Time coincidence between channels $a$ and $b$ for each event, $n$, is investigated using the difference between each \ith peak time in channel $a$ ($t_{n,a,i}$) and each \jth peak time in channel $b$ ($t_{n,b,j}$). We measure the accidental coincidence baseline by comparing times from mismatched events using $t_{n,a,i}$ and $t_{n-1,b,j}$. These accidental coincidence histograms ($H_{\text{shuffled}}(\Delta t; [a,b])$) are referred to as ``shuffled'', as opposed to the ``matched'' histograms ($H_{\text{matched}}(\Delta t; [a,b])$) that contain correlated coincidence. We require that $a$ is less than $b$ to avoid double counting. The matched and shuffled time differences are stored in histograms with bin size of \SI{100}{ps}. If there are two nearby peaks in the same channel, an external coincidence from one peak may appear to also be coincident to the other peak, and that single coincidence may be double-counted. To avoid this mismatch between nearby peaks, we disregard any event $n$ for which the minimum distance between peaks within channel $a$, channel $b$, or channel $b$, event $n-1$ is less than \SI{8}{ns}. We have found that the additional single-electron width in SiPMs read out by channels 8 and 16 on the SCEMAs worsens the timing resolution for those SiPMs such that this analysis is not useful. These channels are therefore not included in the eXT analysis.

The preliminary data shown in \fig{fig:dcprelim} predicts a total eXT probability of \SI{25}{\percent}. Accounting for solid angle, this would lead to a naive channel-to-channel eXT probability of \SI{0.06}{\percent} in pixels directly opposite each other. Given the average DCR of \SI{4.3}{MHz}, the predicted number of eXT photons emitted from channel~$a$ and detected in channel~$b$ is 63. This number is too small to be useful in characterizing the timing or quantity of eXT. In order to bolster our statistics, we take the average probability histogram of all the channels on selected DRS4 chips. There are two DRS4 chips on each SCEMA that each read out a $2\times4$ section of the $2\times8$ array. These groups of $2\times4$ SiPMs are ideal for summing together because they are compact in geometry and should have identical timing. The DRS4 chip number is equal to:
\begin{equation}
\text{chip \#} = \text{Floor}(\text{channel \#}/ 9),
\end{equation}
using the channel number mapping as shown in \fig{fig:channellayout}.

\begin{figure}
\centering 
\includegraphics[width=.99\textwidth]{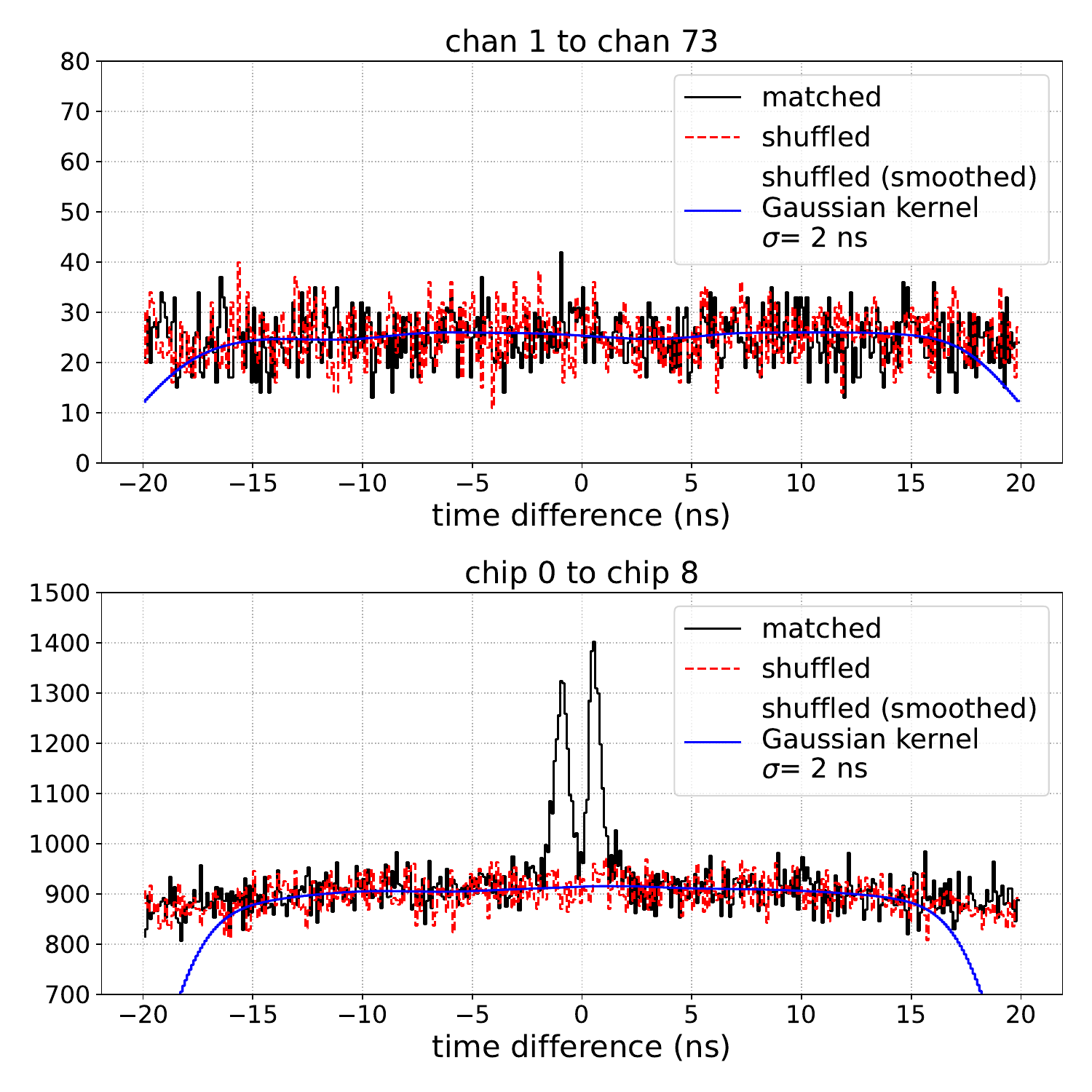}
\caption{\label{fig:baseline_corr} (Top) Example histogram of the peak time differences between channels~1 and~73. The matched (black) and shuffled (red) histograms are described in the text. The blue trace is result of smoothing the shuffled histogram with a \SI{2}{ns} Gaussian kernel. (Bottom) Summed histogram for all channel pairs with channel $a$ on DRS4 chip number~0 and channel~$b$ on DRS4 chip number~8.}
\end{figure}

Examples of the resulting channel-level and chip-level histograms are shown in \fig{fig:baseline_corr}. The matched histogram for coincidence from channel~1 to channel~73 has no clear excess above the shuffled histogram. When the histograms for the rest of the channels on chip~0 and chip~8 are summed together, there are two peaks that emerge. These peaks are approximately symmetrical around zero and are taken to be the ``forward'' and ``backward'' eXT peaks. The peak towards negative time differences represents ``backward'' eXT, which is emitted from SiPMs on chip~8 and detected by SiPMs on chip~0, and the peak at positive time differences represents ``forward'' eXT, which is emitted from SiPMs on chip~0 and detected by SiPMs on chip~8. This convention comes from the requirement that channel~$a$ is less than channel~$b$. 

The shuffled histograms are smoothed using a \SI{2}{ns} Gaussian kernel as is shown in \fig{fig:baseline_corr} to reduce the noise in our baseline calculation and are then subtracted from the matched histograms. The resulting baseline-subtracted histograms should only contain external crosstalk. For a single pair of channels, $(a, \ b)$, the time-dependent eXT probability density is defined as:
\begin{equation} \label{eq:pext_hist}
\peXT(\Delta t; [a,b]) = \left( H_{\text{matched}}(\Delta t; [a,b]) - \texttt{smooth}(H_{\text{shuffled}}(\Delta t; [a,b])) \right)/N_a,
\end{equation}
where $N_a$ is the total number of peaks from channel~$a$ used to generate the $\Delta t$ histograms. This formulation for $\peXT(\Delta t; [a,b])$ is only valid for forward~eXT, since $N_a$ is used as the normalization. For channel pairs $[b,a]$, where $b>a$, the shuffled and matched histograms are first flipped around $\Delta t = 0$:
\begin{equation} \label{eq:pext_hist_bac}
\peXT(\Delta t; [b,a]) = \left( H_{\text{matched}}(-\Delta t; [a,b]) - \texttt{smooth}(H_{\text{shuffled}}(-\Delta t; [a,b])) \right)/N_b.
\end{equation}

\begin{figure}
\centering 
\includegraphics[width=.99\textwidth]{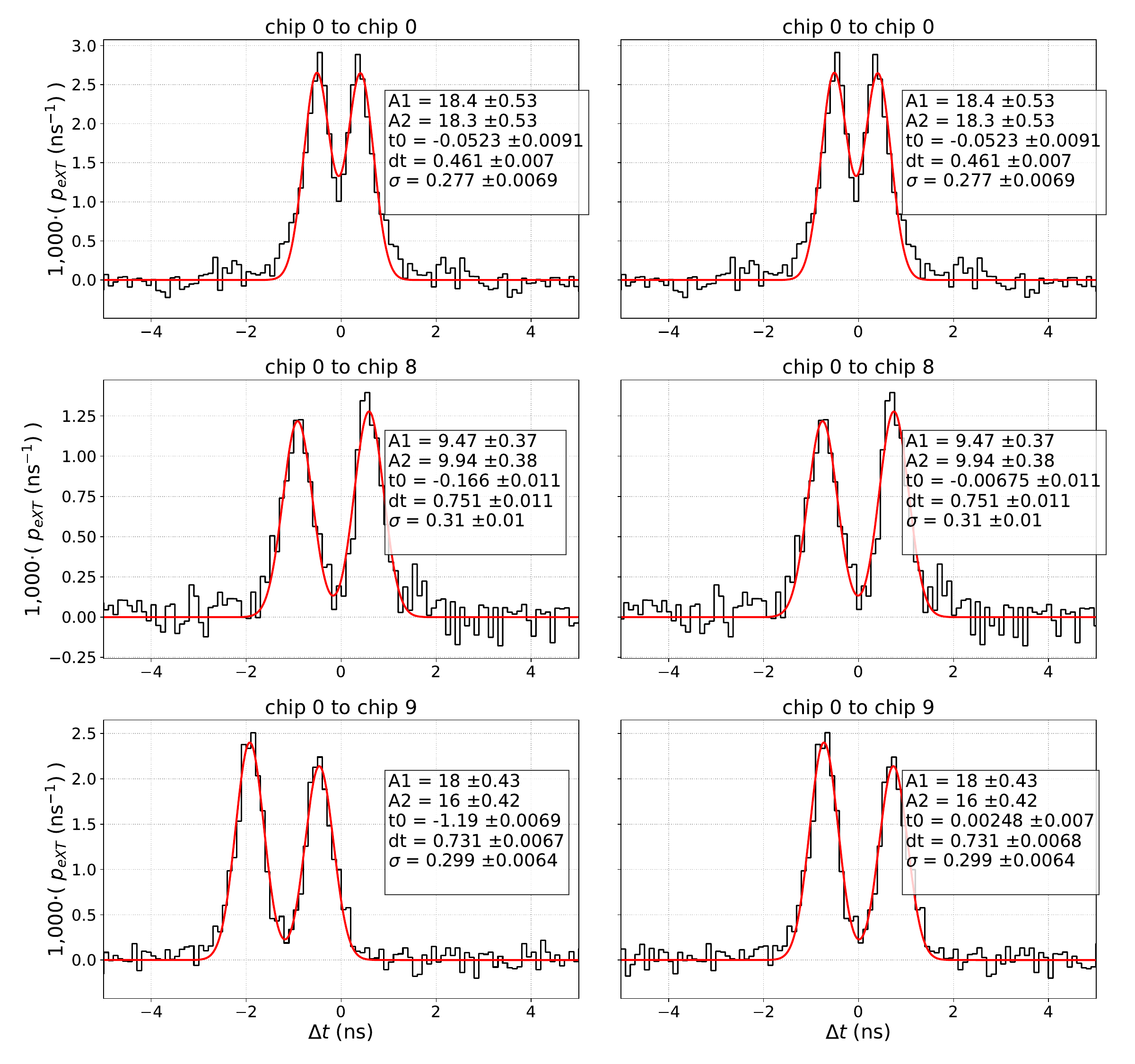}
\caption{\label{fig:chip2chip} Examples of average time-dependent channel-to-channel eXT probability density for chips (0, 0), chips (0,8), and chips (0,9). The probability density is in units of eXT photons per ns per 1,000 primary photons. Optimizations of the fitting model described in \eqnref{eq:extfit} are shown in red. The optimized parameters and fitting uncertainties are reported in the text boxes. The left-hand column shows the histograms with raw times, and the right hand column shows the histograms after accounting for offsets in intra-board DRS4 timing as described in the text.    }
\end{figure}

\Fig{fig:chip2chip} shows examples of the average probability density for all channel-pairs, $[a,b]$ with channel~$a$ on chip $A$ and channel~$b$ on chip~$B$. The chip pairs $\{A,B\}$ shown in \fig{fig:chip2chip} are $\{0,0\}$, $\{0,8\}$, and $\{0,9\}$. Forward and backward eXT peaks are clearly visible in all of these chip pairs. We find the area and position of these peaks by fitting the sum of two Gaussian functions to the measured spectrum:
\begin{equation}\label{eq:extfit}
y = \frac{A_1}{\sqrt{2\pi \sigma^2}}e^{-(t-(t_0-dt))^2/(2\sigma^2)}+\frac{A_2}{\sqrt{2\pi \sigma^2}}e^{-(t-(t_0+dt))^2/(2\sigma^2)}
\end{equation}
The fitting parameters $A_1$ and $A_2$ correspond to the areas of the backward and forward eXT peaks, respectively. 
The parameter $A_2$ is defined as the measured total forward eXT probability, $P_{\text{eXT},[a,b]}$. 
The parameter $A_1$ is related to the total backward eXT probability, but is not properly normalized.
The two peaks are forced to share a width, $\sigma$, which has contributions from single-photon timing resolution, variations in the emission times of eXT photons, and variation in photon travel time between the various channel pairs. If the digitizer clocks are properly aligned, the central time, $t_0$, should be \SI{0}{ns}. The fitting parameter $dt$ corresponds to the average time between the primary photon being detected in channel~$a$ and the eXT photon being detected in channel~$b$.

We find that, for even-to-odd chip pairs, $t_0$ is offset by more than \SI{-1}{ns}, and for odd-to-even chip pairs, it is offset by more than \SI{+1}{ns}. This cannot be physical, because it would require that the eXT photon arrived at the secondary pixel before the primary photon was detected. We believe this offset is due to a misalignment between the clocks on the two DRS4 chips on each SCEMA. We measure $t_0$ for each chip pair $\{3,B\}$, for $B = (8:15)$, and  $\{A,10\}$, for $A = (0:7)$. The time-axis of the histograms are then adjusted such that chips $(0:7)$ are aligned with chip~10, and chips $(8:15)$ are aligned with chip 3. Chip~10 is made to align with chip~3, so that all chips should share a common alignment. For each chip pair, we are left with a realignment value, $\Delta t_{0, \{A,B\}}$. The set of time-separation values, $\Delta t$, corresponding to the binning used to create the coincidence histograms is transformed according to:
\begin{equation}
\peXT(\Delta t; [a,b]) \rightarrow \peXT(\Delta t - \Delta t_{0, \{A,B\}}; [a,b])
\end{equation}
The left-hand column of \fig{fig:chip2chip} shows examples the average \peXT for several chip pairs prior to this adjustment, and the right-hand column shows the result of the alignment.

\begin{figure}
\centering 
\includegraphics[width=.85\textwidth]{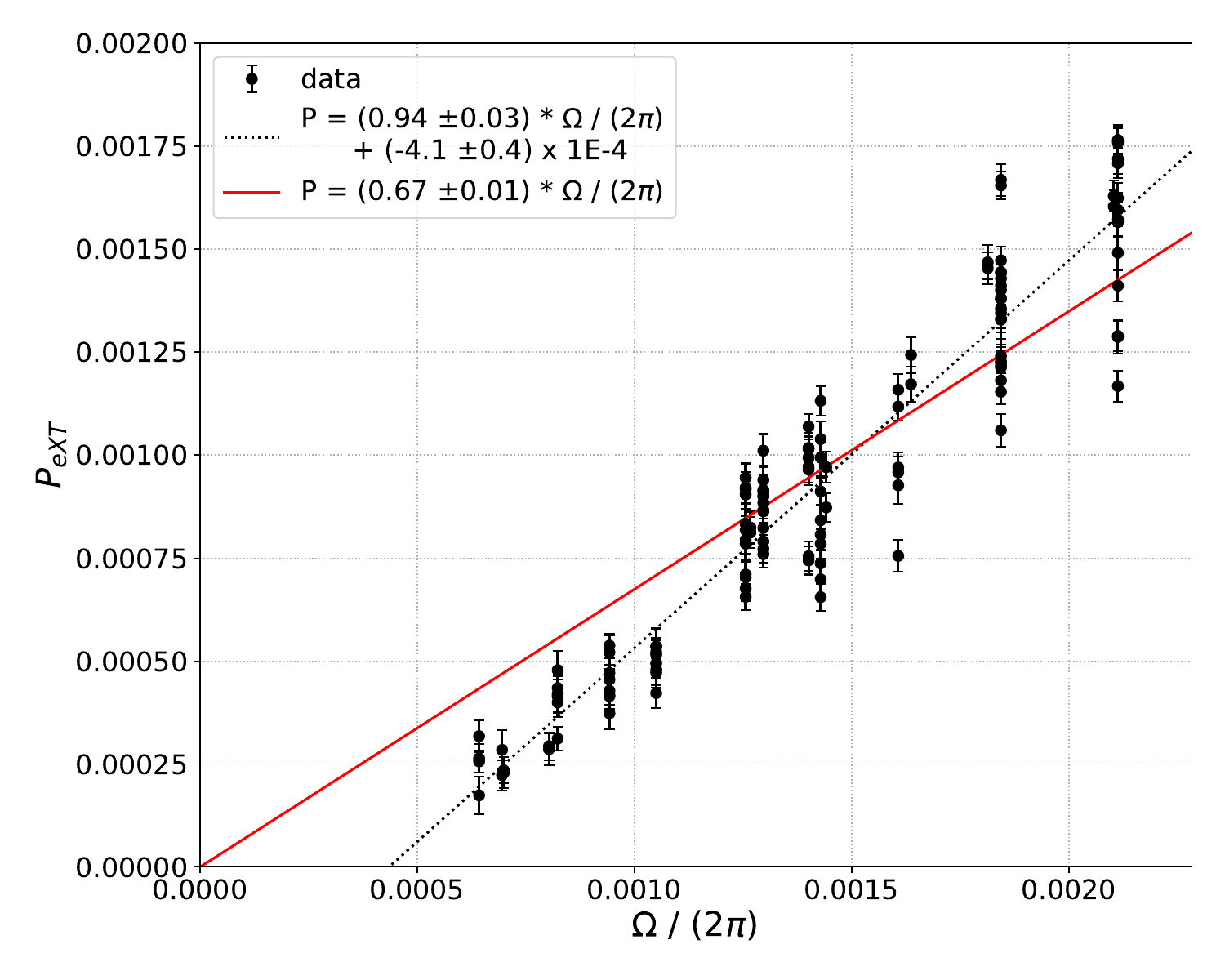}
\caption{\label{fig:pvsa} Average eXT probability as a function of solid angle efficiency for all chip pairs in which chip A and chip B are in opposite planes. The data contains both forward and backward channel pairs. The solid red line shows a linear fit where the x-intercept is defined to be 0, as would be expected if both the PDE and emission of eXT photons were isotropic. The dashed line shows a linear fit that allows the zero-crossing to float. The upper limit of the x-scale is set to be equal to the angular efficiency of a pair of SiPMs directly opposite each other, which is the maximum possible angular efficiency.
}
\end{figure}

The average $\peXT(\Delta t - \Delta t_{0, \{A,B\}}; [a,b])$ and $p_{\text{eXT},[a,b]}$ is calculated for all chip-pairs, $\{A,B\}$ such that $A$ and $B$ are on opposite sides of the detector. 
In \fig{fig:pvsa}, the average \peXT between these chips is plotted against the average geometric efficiency between SiPMs. Since eXT photons are emitted away from the primary SiPM and into the detector volume, we define the angular efficiency to be equal to the solid angle divided by $2\pi$. \Fig{fig:pvsa} shows that the eXT probability increases with solid angle as expected, but the relationship is not one of pure proportionality. We do not find a significant difference between forward and backward eXT. The best fit line has a zero-crossing well before the geometric efficiency drops to zero. This suggests an angular dependence to either the eXT photon emission, the SiPM PDE, or both. We calculate an average effective total eXT probability by tiling a plane with SiPM-sized squares and summing the eXT probability for each square given its solid angle. These probabilities, $p$, are calculated using the black dashed line in \Fig{fig:pvsa}, where the condition $p \geq 0$ is enforced. We find the average effective total eXT probability is approximately \SI{22}{\percent}.

\begin{figure}
\centering 
\includegraphics[width=.85\textwidth]{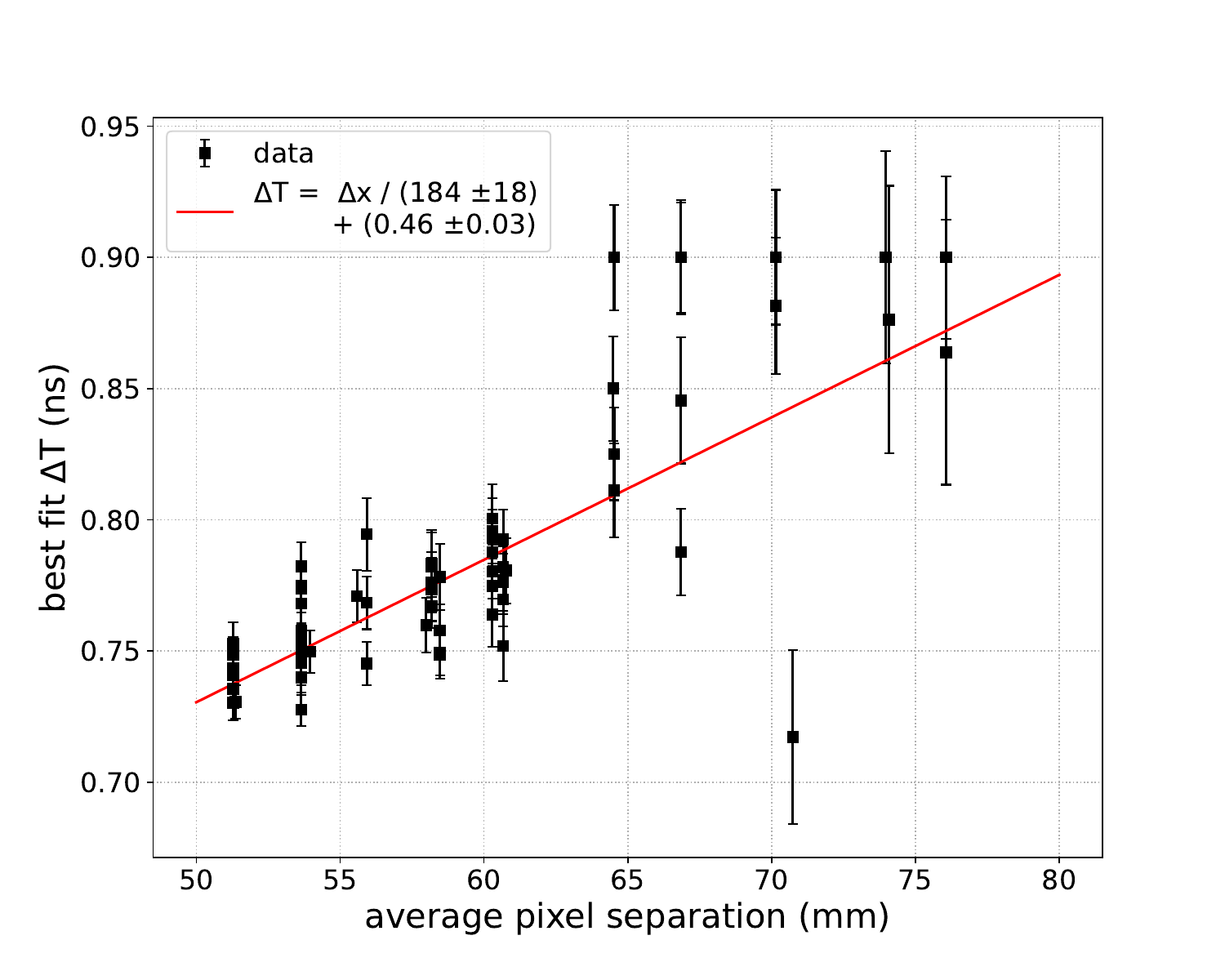}
\caption{\label{fig:dtvdx} Best fit time offset as a function of average pixel separation for all chip pairs in which chip A and chip B are in opposite planes. The solid red line shows a linear fit where both the slope and intercept are allowed to vary. The slopes are reported in units of mm/ns.
}
\end{figure}

\Fig{fig:dtvdx} shows the optimized $dt$ for all opposite-array chip-pairs, plotted against the average SiPM separation. A linear fit to these data shows good agreement with a slope equal to the speed of light in EJ-204 (\SI{190}{mm/ns}). These fits also suggest that even at a SiPM separation of \SI{0}{cm}, there will be about a \SI{0.5}{ns} delay between the primary photon detection, and the eXT photon detection. It is not clear what causes this zero-separation delay. 

We use the optimized $\sigma$ values for all opposite-array chip-pairs to constrain the in-situ timing resolution of the SiPMs. These widths range from \SI{277(11)}{ps} to \SI{354(49)}{ps}, with an average value of \SI{307(2)}{ps}. It is not clear exactly how these widths relate to the SPTR of the SiPMs. There will be contributions to $\sigma$ from the uncertainty in the primary photon detection time, the uncertainty in the emission time of the eXT photon, differences in path length between the various SiPMs in the chip-pair, and the uncertainty in the eXT photon detection time. We obtain an approximate upper limit on photon time uncertainty by assuming that $\sigma$ is dominated by the SPTR of the photon detections and that SPTR for the primary and eXT photons are identical. Under these assumptions, we find that the upper limit of the photon time uncertainty for SiPMs in the detector is \SI{250(35)}{ps}. 

The eXT between SiPMs on the same side of the detector is expected to be small, since there is no direct path for eXT photons. Any photons exchanged between these channels must go through at least one reflection, so eXT will be suppressed. The intra-array eXT is typically too small to be calculated using the chip-to-chip method described above. Instead, we take the average of $\peXT(\Delta t; [a,b])$ for each channel $a$ in one of the $8\times8$ arrays to each channel $b>a$ in the same $8\times8$ array. The individual normalization for $\peXT(\Delta t; [a,b])$ is modified so that negative time bins are divided by $N_b$, and positive time bins are divided by $N_a$. The average channel-to-channel probability for intra-array eXT is \SI[separate-uncertainty = true]{0.033(0.001)}{\percent} for the array located at z = \SI{+25}{mm} and \SI[separate-uncertainty = true]{0.029(0.001)}{\percent} for the array located at z = \SI{-25}{mm}.

\begin{figure}
\centering 
\includegraphics[width=.99\textwidth]{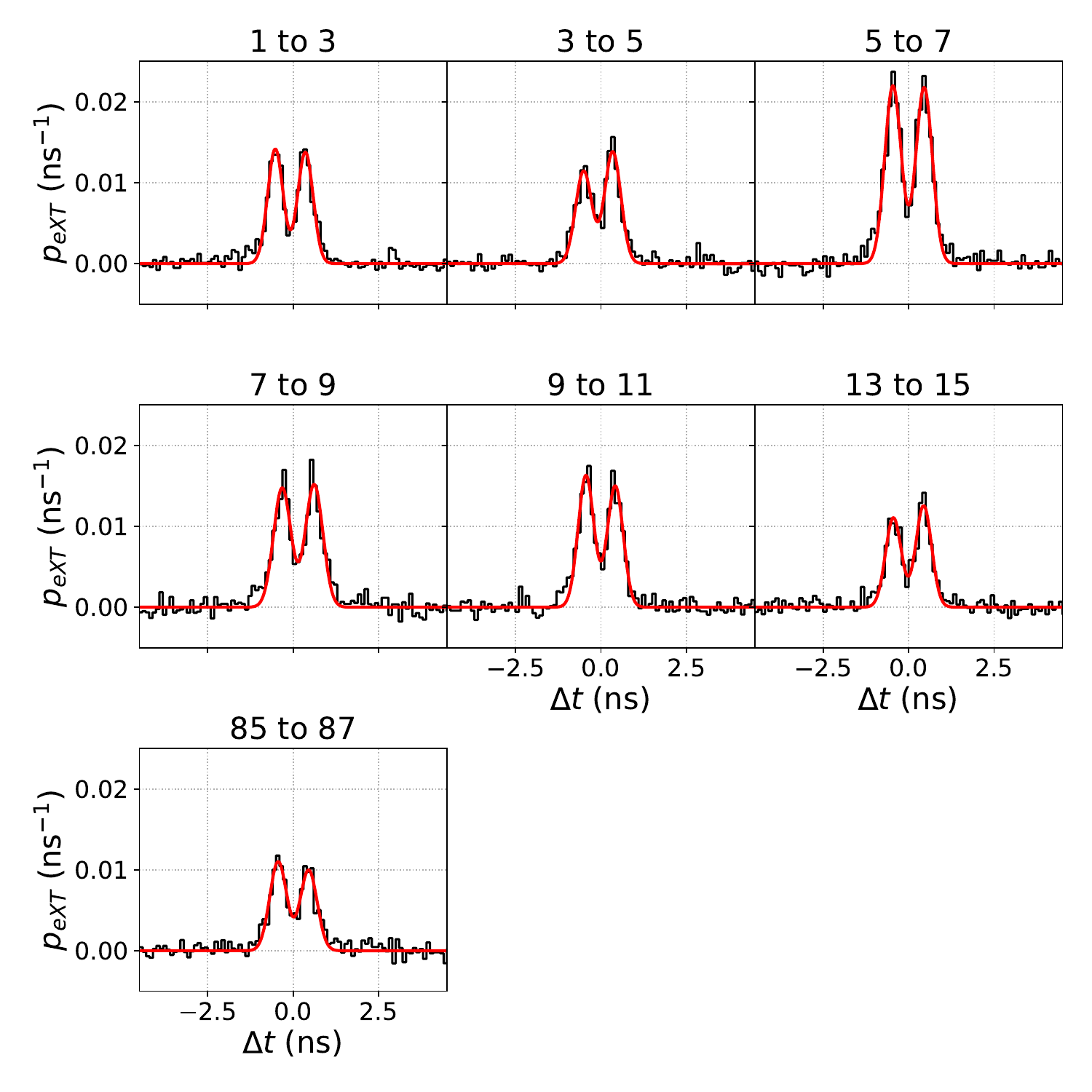}
\caption{\label{fig:highpairs} Channel-pairs, $[a,b]$, for which individual eXT probability is greater than \SI{1}{\percent}. The probability at negative times is normalized to $N_b$, and the probability at positive times is normalized to $N_a$. This means that both the forward and backward probability peaks are correctly normalized. The red curves show the optimizations for \eqnref{eq:extfit}. }
\end{figure}

There are several individual channel pairs with eXT probability greater than \SI{1}{\percent}, as is shown in \fig{fig:highpairs}. These pairs are all nearest neighbors and are on the edge of the scintillator. This may indicate bubbles in the optical coupling layer that create a reflection path for eXT photons to be exchanged.

\begin{figure}
\centering 
\includegraphics[width=.85\textwidth]{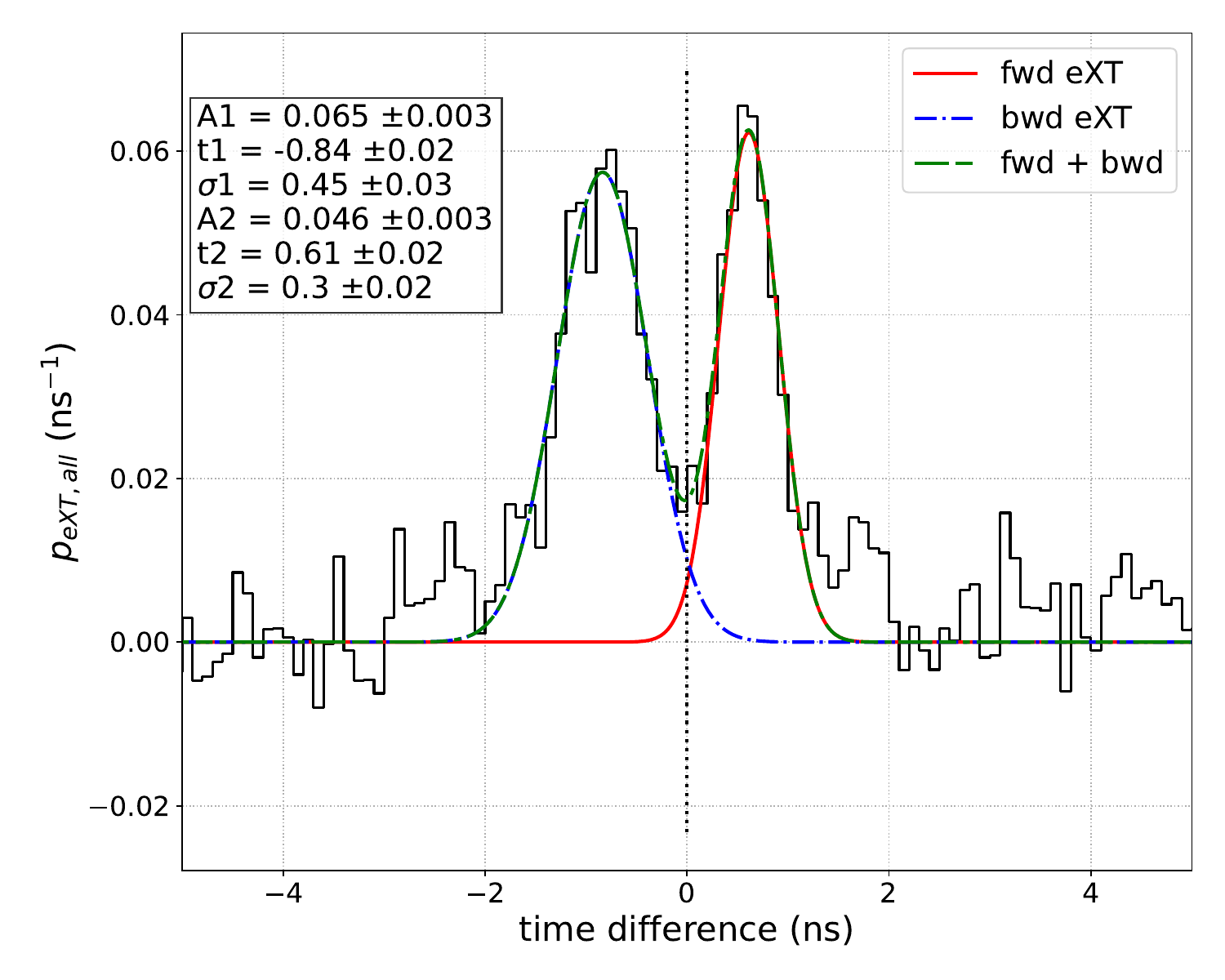}
\caption{\label{fig:ch2all} Example of the summed channel-to-all eXT probability density for SiPM channel 1.}
\end{figure}

For each SiPM channel, $a$, in the detector, we calculate the total probability that an eXT photon generated in channel $a$ will be detected in any other channel. This channel-to-all eXT probability density is calculated by summing together $\peXT(\Delta t - \Delta t_{0, \{A,B\}}; [a,b])$ for all $b \neq a$. 
Each chip-pair will have a different $\Delta t_{0, \{A,B\}}$ so effectively has a different set of sample times for time-separation. The channel-to-all probabilities are calculated over multiple DRS chip-pairs, so the measured values of $\peXT(\Delta t - \Delta t_{0, \{A,B\}}; [a,b])$ are passed through linear interpolation to produce a new set of probability densities ($\tilde{p}_{\text{eXT}}(\Delta t; [a,b])$) that are evaluated at the original values of $\Delta t$.
We sum together $\tilde{p}_{\text{eXT}}(\Delta t; [a,b])$ for all $b \neq a$ to calculate the channel-to-all probability density:
\begin{equation}
\peXTall(\Delta t; a) = \sum_{b \neq a}\tilde{p}_{\text{eXT}}(\Delta t; [a,b])
\end{equation}
An example of $\peXTall(\Delta t; a)$ for $a=1$ is shown in \fig{fig:ch2all}. A double-Gaussian fitting model similar to \eqnref{eq:extfit} is again applied. The equation is adjusted such that the widths and locations of the two Gaussians are floated independently. The area of the second peak is reported as the summed channel-to-all eXT probability ($\PeXTall(a)$).

\begin{figure}
\centering 
\includegraphics[width=.99\textwidth]{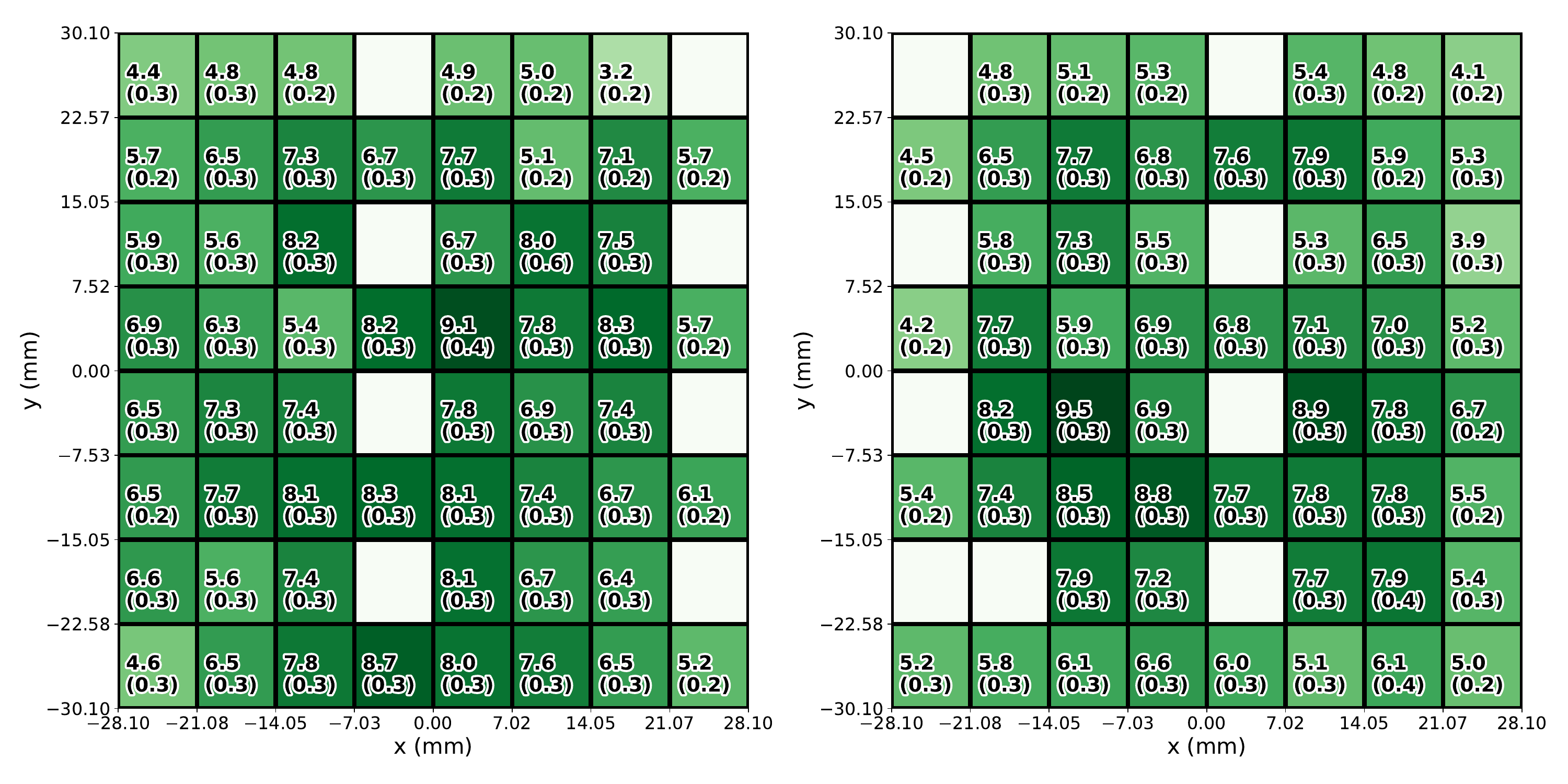}
\caption{\label{fig:extmap} The summed channel-to-all eXT for each SiPM channel. Channel mapping is shown in \fig{fig:channellayout}. Channels 8 and 16 are on each SCEMA are not included in the analysis. Channel 86 is non-functional. }
\end{figure}

\Fig{fig:extmap} shows the measurements of $\PeXTall(a)$ for all SiPM channels in the detector. As noted above, channels 8 and 16 are on each SCEMA are not included in the analysis. The SiPM channels located near the edge of the detector tend to have the smallest \PeXTall. This is expected from solid angle considerations since they are nearest to the black-painted sides of the scintillator.  

\section{Summary}
We report on the design and commissioning of a prototype monolithic kinematic neutron imaging detector. This prototype employs a single monolithic volume of EJ-204 fast plastic scintillator coupled to Hamamatsu S13360-6075PE MPPCs (SiPMs) in place of more traditional photon detectors such as PMTs. These SiPMs are mounted on and read out using analog electronics that are optimized for the measurement of single photons and reduction of electronic crosstalk. The signals are digitized at approximately \SI{5}{GHz} using custom-built electronics boards that employ switched capacitor array waveform sampling provided by DRS4 chips. 

The impact on event reconstruction due to optical crosstalk and dark counts from the SiPMs is described and evaluated in section~\ref{sec:likelihood}. An updated likelihood model is presented that accounts for these effects. We test this likelihood model using a simple Monte-Carlo simulation and find that it is able to reconstruct unbiased event parameters. This model still needs to be exercised and validated on single-site scintillation data before it is employed for multi-site events and image reconstruction.

In section \ref{sec:sptr}, we describe a method of single-photon analysis capable of measuring photon arrival times with a maximum precision of \SI[separate-uncertainty = true]{121(3)}{ps} by employing a template-fitting method on data taken using a Photek LPG-405 fast-pulsed laser. A faster analysis that uses a linear fit to the rising edge is found to have a slightly worse precision of \SI[separate-uncertainty = true]{128(4)}{ps}. 

In section~\ref{sec:darkcounts}, we measure the single-electron response pulse-height spectrum for each of the 128 SiPM channels from dark counts. We apply a fitting model to the pulse-height spectrum to calculate the DCR, iXT, single-electron gain, and single electron width for all 128 SiPMs without the need for external calibration. These dark count calibration measurements revealed several position-dependent effects that were not anticipated prior to assembly. In particular, we observe elevated DCR and iXT in SiPMs at the +y edge of the detector. This effect was was not observed in testing prior to assembly, and we do not have a complete understanding of its origin. We also measure the effect of temperature on the predicted gain resolution by comparing the variations in a subset of the data to the variations in the entire data-set. These effects will impact the accuracy of the likelihood reconstruction described in section~\ref{sec:likelihood}, so it is important to perform these real-time, in-situ calibrations.

Finally, we perform a time-correlation analysis to measure the probability and time profile of the eXT photons exchanged between SiPM channels in section~\ref{sec:timecorr}. Using this analysis, we confirm a single-photon time precision of less than \SI{250(34)}{ps} in the assembled detector. We also find a misalignment between the DRS4 clocks and correct for it using the measured time correlations. We find an unexplained delay in the eXT of, on average, \SI{460 (30)}{ps} at \SI{0}{mm} SiPM separation. This delay may be due to the finite difference between photon incidence time and the time of maximum avalanche current within the SiPMs, but further investigation is required. This delay is equivalent to the $t_a$ shown in equations~\ref{eq:ctprof_1} and~\ref{eq:ctprof_m} and is important to characterize in order to produce unbiased event reconstruction.

\acknowledgments
Sandia National Laboratories is a multimission laboratory managed and 
operated by National Technology \& Engineering Solutions of Sandia, LLC, a 
wholly owned subsidiary of Honeywell International Inc., for the U.S.\ 
Department of Energy’s National Nuclear Security Administration under 
contract DE-NA0003525.

This work was performed in part under the auspices of the U.S. Department of Energy 
by Lawrence Berkeley National Laboratory under Contract DE-AC02-05CH11231. 
The project was funded by the U.S. Department of Energy, National Nuclear Security Administration, 
Office of Defense Nuclear Nonproliferation Research and Development (DNN R\&D). 

This paper describes objective technical results and analysis. Any subjective 
views or opinions that might be expressed in the paper do not necessarily 
represent the views of the U.S.\ Department of Energy or the United States 
Government.

This article has been authored by an employee of National Technology \& 
Engineering Solutions of Sandia, LLC under Contract No. DE-NA0003525 with the 
U.S.\ Department of Energy (DOE). The employee owns all right, title and 
interest in and to the article and is solely responsible for its contents. 
The United States Government retains and the publisher, by accepting the 
article for publication, acknowledges that the United States Government 
retains a non-exclusive, paid-up, irrevocable, world-wide license to publish 
or reproduce the published form of this article or allow others to do so, for 
United States Government purposes. The DOE will provide public access to 
these results of federally sponsored research in accordance with the DOE 
Public Access Plan
\url{https://www.energy.gov/downloads/doe-public-access-plan}.

Document Release Number: SAND2023-09786O

\bibliography{SVSC_2ndMono_Paper}

\providecommand{\href}[2]{#2}\begingroup\raggedright\begin{thebibliography}{10}

\bibitem{Vanier2007}
P.E.~Vanier, L.~Forman, I.~Dioszegi, C.~Salwen and V.J.~Ghosh,
  \emph{Calibration and testing of a large-area fast-neutron directional
  detector},  in \emph{2007 IEEE Nuclear Science Symposium Conference Record},
  vol.~1, pp.~179--184, 2007,
  \href{https://doi.org/10.1109/NSSMIC.2007.4436312}{DOI}.

\bibitem{Bravar2009}
U.~Bravar, R.S.~Woolf, P.J.~Bruillard, E.O.~Fluckiger, J.S.~Legere,
  A.L.~MacKinnon et~al., \emph{Calibration of the fast neutron imaging
  telescope ({FNIT}) prototype detector},
  \href{https://doi.org/10.1109/TNS.2009.2028025}{\emph{IEEE Transactions on
  Nuclear Science} {\bfseries 56} (2009) 2947}.

\bibitem{Brennan2011}
J.~Brennan, E.~Brubaker, R.~Cooper, M.~Gerling, C.~Greenberg, P.~Marleau
  et~al., \emph{Measurement of the fast neutron energy spectrum of an
  $^{241}\rm{Am\!-\!Be}$ source using a neutron scatter camera},
  \href{https://doi.org/10.1109/TNS.2011.2163192}{\emph{IEEE Transactions on
  Nuclear Science} {\bfseries 58} (2011) 2426}.

\bibitem{Goldsmith2016}
J.E.M.~Goldsmith, M.D.~Gerling and J.S.~Brennan, \emph{A compact neutron
  scatter camera for field deployment},
  \href{https://doi.org/10.1063/1.4961111}{\emph{Review of Scientific
  Instruments} {\bfseries 87} (2016) 083307}
  [\href{https://arxiv.org/abs/https://doi.org/10.1063/1.4961111}{{\ttfamily
  https://doi.org/10.1063/1.4961111}}].

\bibitem{braverman2018}
J.~Braverman, J.~Brennan, E.~Brubaker, B.~Cabrera-Palmer, S.~Czyz, P.~Marleau
  et~al., \emph{Single-volume neutron scatter camera for high-efficiency
  neutron imaging and spectroscopy},  2018.

\bibitem{keefe2022}
K.~Keefe, H.~Alhajaji, E.~Brubaker, A.~Druetzler, A.~Galindo-Tellez, J.~Learned
  et~al., \emph{Design and characterization of an optically segmented single
  volume scatter camera module},
  \href{https://doi.org/10.1109/tns.2022.3172002}{\emph{{IEEE} Transactions on
  Nuclear Science} {\bfseries 69} (2022) 1267}.

\bibitem{ej232q_datasheet}
E.~Technology, ``{EJ-232Q Data Sheet}.''

\bibitem{ej204_datasheet}
{Eljen Technology}, ``General purpose plastic scintillator {EJ-200}, {EJ-204},
  {EJ-208}, {EJ-212}.''
  \url{https://eljentechnology.com/images/products/data_sheets/EJ-200_EJ-204_EJ-208_EJ-212.pdf},
  July, 2021.

\bibitem{s13360_datasheet}
{Hamamatsu Photonics}, ``{S13360} series: {MPPCs} for precision measurement.''
  \url{https://www.hamamatsu.com/resources/pdf/ssd/s13360_series_kapd1052e.pdf},
  Aug., 2016.

\bibitem{acerbi2019}
F.~Acerbi and S.~Gundacker, \emph{Understanding and simulating {SiPMs}},
  \href{https://doi.org/https://doi.org/10.1016/j.nima.2018.11.118}{\emph{Nuclear
  Instruments and Methods in Physics Research Section A: Accelerators,
  Spectrometers, Detectors and Associated Equipment} {\bfseries 926} (2019)
  16}.

\bibitem{cates2022}
J.W.~Cates, J.~Steele, J.~Balajthy, V.~Negut, P.~Hausladen, K.~Ziock et~al.,
  \emph{Front-end design for {SiPM}-based monolithic neutron double scatter
  imagers}, \href{https://doi.org/10.3390/s22093553}{\emph{Sensors} {\bfseries
  22} (2022) }.

\bibitem{ej510_datasheet}
{Eljen Technology}, ``Reflective paint {EJ-510}.''
  \url{https://eljentechnology.com/images/products/data_sheets/EJ-510.pdf},
  July, 2021.

\bibitem{ej560_datasheet}
{Eljen Technology}, ``Silicone rubber optical interface {EJ-560}.''
  \url{https://eljentechnology.com/images/products/data_sheets/EJ-560.pdf},
  July, 2021.

\bibitem{trippliteAC_datasheet}
{Tripp Lite by Eaton}, ``Portable {AC} unit for server rooms - 12,000 {BTU}
  (3.5 {kW}), {120V}.''
  \url{https://assets.tripplite.com/product-pdfs/en/srcool12k.pdf}, 2022.

\bibitem{dg535_datasheet}
{Stanfor Research Systems}, ``Model {DG535} digital delay / pulse generator.''
  \url{https://www.thinksrs.com/downloads/pdfs/manuals/DG535m.pdf}, Apr., 2017.

\bibitem{steele2019}
J.~Steele, J.~Brown, E.~Brubaker and K.~Nishimura, \emph{{SCEMA}: a high
  channel density electronics module for fast waveform capture},
  \href{https://doi.org/10.1088/1748-0221/14/02/p02031}{\emph{Journal of
  Instrumentation} {\bfseries 14} (2019) P02031}.

\bibitem{ritt2010}
S.~Ritt, R.~Dinapoli and U.~Hartmann, \emph{Application of the {DRS} chip for
  fast waveform digitizing},
  \href{https://doi.org/10.1016/j.nima.2010.03.045}{\emph{Nuclear Instruments
  and Methods in Physics Research Section A: Accelerators, Spectrometers,
  Detectors and Associated Equipment} {\bfseries 623} (2010) 486}.

\bibitem{lpg405}
{Photek}, ``{LPG-405} pulsed laser for time-resolved detector diagnostics.''
  \url{https://www.photek.com/pdf/datasheets/electronics/DS036-LPG-405-Datasheet-issue-2.pdf},
  Dec., 2017.

\bibitem{Laplace2020}
T.~Laplace, B.~Goldblum, J.~Brown, D.~Bleuel, C.~Brand, G.~Gabella et~al.,
  \emph{Low energy light yield of fast plastic scintillators},
  \href{https://doi.org/https://doi.org/10.1016/j.nima.2018.10.122}{\emph{Nuclear
  Instruments and Methods in Physics Research Section A: Accelerators,
  Spectrometers, Detectors and Associated Equipment} {\bfseries 954} (2020)
  161444}.

\bibitem{Masuda2021}
T.~Masuda, D.G.~Ang, N.R.~Hutzler, C.~Meisenhelder, N.~Sasao, S.~Uetake et~al.,
  \emph{Suppression of the optical crosstalk in a multi-channel silicon
  photomultiplier array}, \href{https://doi.org/10.1364/oe.424460}{\emph{Optics
  Express} {\bfseries 29} (2021) 16914}.

\bibitem{biland2014}
A.~Biland, T.~Bretz, J.~Bu{\ss}, V.~Commichau, L.~Djambazov, D.~Dorner et~al.,
  \emph{Calibration and performance of the photon sensor response of {FACT}
  {\textemdash} the first g-{APD} cherenkov telescope},
  \href{https://doi.org/10.1088/1748-0221/9/10/p10012}{\emph{Journal of
  Instrumentation} {\bfseries 9} (2014) P10012}.

\bibitem{Barlow_1990}
R.~Barlow, \emph{Extended maximum likelihood},
  \href{https://doi.org/https://doi.org/10.1016/0168-9002(90)91334-8}{\emph{Nuclear
  Instruments and Methods in Physics Research Section A: Accelerators,
  Spectrometers, Detectors and Associated Equipment} {\bfseries 297} (1990) 496
  }.

\bibitem{scipy}
P.~Virtanen, R.~Gommers, T.E.~Oliphant, M.~Haberland, T.~Reddy, D.~Cournapeau
  et~al., \emph{{{SciPy} 1.0: Fundamental Algorithms for Scientific Computing
  in Python}}, \href{https://doi.org/10.1038/s41592-019-0686-2}{\emph{Nature
  Methods} {\bfseries 17} (2020) 261}.

\end{thebibliography}\endgroup

\clearpage

\appendix

\section{Characterization Parameters}
\label{app:parms}

In this appendix, we list all of the parameters required for the likelihood analysis described in \secref{sec:likelihood}. We also describe the method of characterization for each parameter.

\begin{table}[htbp!]
\begin{center}
\caption{List of detector parameters required for event reconstruction.}
\vspace*{1pt}
\label{tab:params}
\begin{tabular}{p{0.114\textwidth}|p{0.37\textwidth}|p{0.43\textwidth}}
\hline
\textbf{Parameter} & \textbf{Description} & \textbf{Measurement}\\
\hline \hline
\Galphax & Geometric efficiency of a photon generated isotropically at position $\vec{x}$ to be detected in pixel~$\alpha$. & Can be approximated using \eqnref{eq:eff_ideal}, but this ignores several effects such as reflections. Can also be measured directly. \\
\hline
$\Omega_{\alpha}(\vec{x})$  & Solid angle subtended by pixel~$\alpha$ from position $\vec{x}$. Needed to calculate \Galphax (\eqnref{eq:eff_ideal}). & Calculated analytically. \\
\hline
$PDE_{\alpha}$ & Photon detection efficiency for pixel~$\alpha$. Needed to calculate \Galphax (\eqnref{eq:eff_ideal}). & Ideally measured in-situ using a calibrated light source. Can be approximated using datasheet. \\
\hline
$\lambda$ & Photon attenuation length. Needed to calculate \Galphax (\eqnref{eq:eff_ideal}). & Taken from scintillator datasheet. \\
\hline
$\mathcal{F}_{\alpha}(t)$ & Photon time distribution for scintillator emission, convolved with pixel~$\alpha$ time response. & Primary emission taken from datasheet or measured using a separate test stand. \\
\hline
$R_{\alpha}$ & Total dark count rate in pixel~$\alpha$. Accounts for contributions from external crosstalk. & Measured directly from dark counts. Equal to the number of peaks in a dark count data-set, divided by the data-set time as described in \secref{sec:darkcounts}.\\
\hline
$\mu_{\alpha}$ & Borel parameter for internal crosstalk in pixel~$\alpha$.  & Taken from the single-photon pulse-height spectrum measured from dark counts as described in \secref{sec:darkcounts}. \\
\hline 
$x_{g,\alpha}$ & Average pulse-height of a single photon in pixel~$\alpha$. & Measured using dark count pulse-height spectrum as described in \secref{sec:darkcounts}.\\
\hline
$c_{\beta \alpha}$ & Pixel-to-pixel external crosstalk. Probability that a crosstalk photon from a single-photon pulse in pixel~$\beta$ will be detected in pixel~$\alpha$. & Measured in-situ in the final detector assembly using time-correlations of dark counts as described in \secref{sec:timecorr}. \\
\hline 
$t_a$ & Avalanche delay. Average time between photon incidence and emission time for external crosstalk. & Measured from time-correlations of dark counts as described in \secref{sec:timecorr}.\\
\hline
$A_{sp,\alpha}$ & Average pulse-area of a single photon. & Measured in-situ using data-set under investigation. Photon peaks are clearly distinguishable in scintillation data.\\
\hline
\end{tabular}
\end{center}
\end{table}

\clearpage

\section{Dark Count Spectrum Fits}
\label{app:dcrfits}

In this appendix we include figures displaying fits to the dark count spectra observed by the SiPMs on SCEMA boards 1--8, as shown for board~0 in \fig{fig:dcrfit0} (\secref{sec:darkcounts}).


\begin{figure}[htbp!]
\centering 
\includegraphics[width=.99\textwidth]{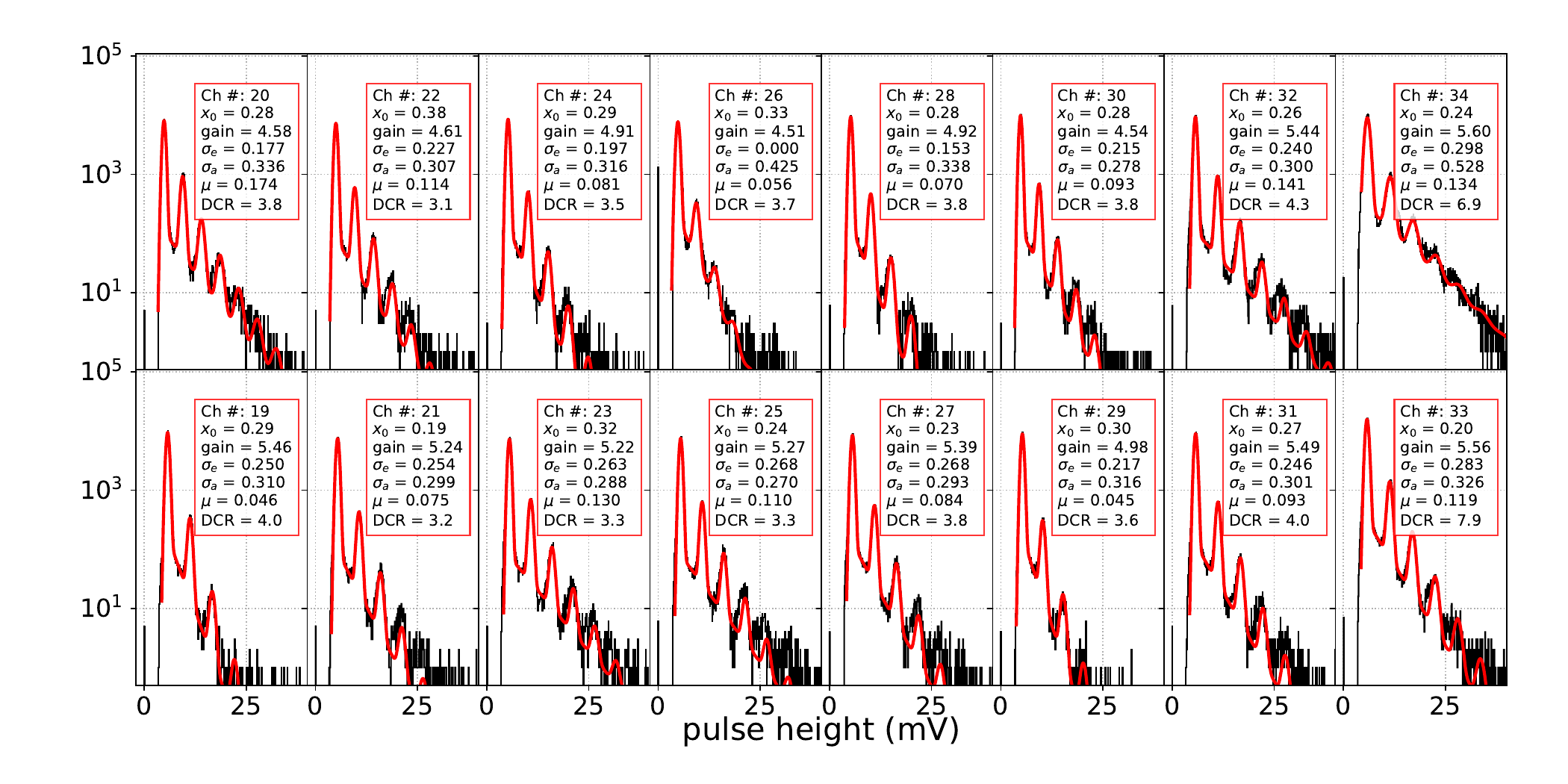}
\caption{\label{fig:dcrfit1} Fits to the pulse-height spectra measured from the dark count data-set for all channels on SCEMA board 1. The fitting model is described in \secref{sec:dcrintro}. The gain has units of filtered-mV per photon, the crosstalk parameter is unitless, and the dark count rate has units of MHz.   }
\end{figure}

\begin{figure}[htbp!]
\centering 
\includegraphics[width=.99\textwidth]{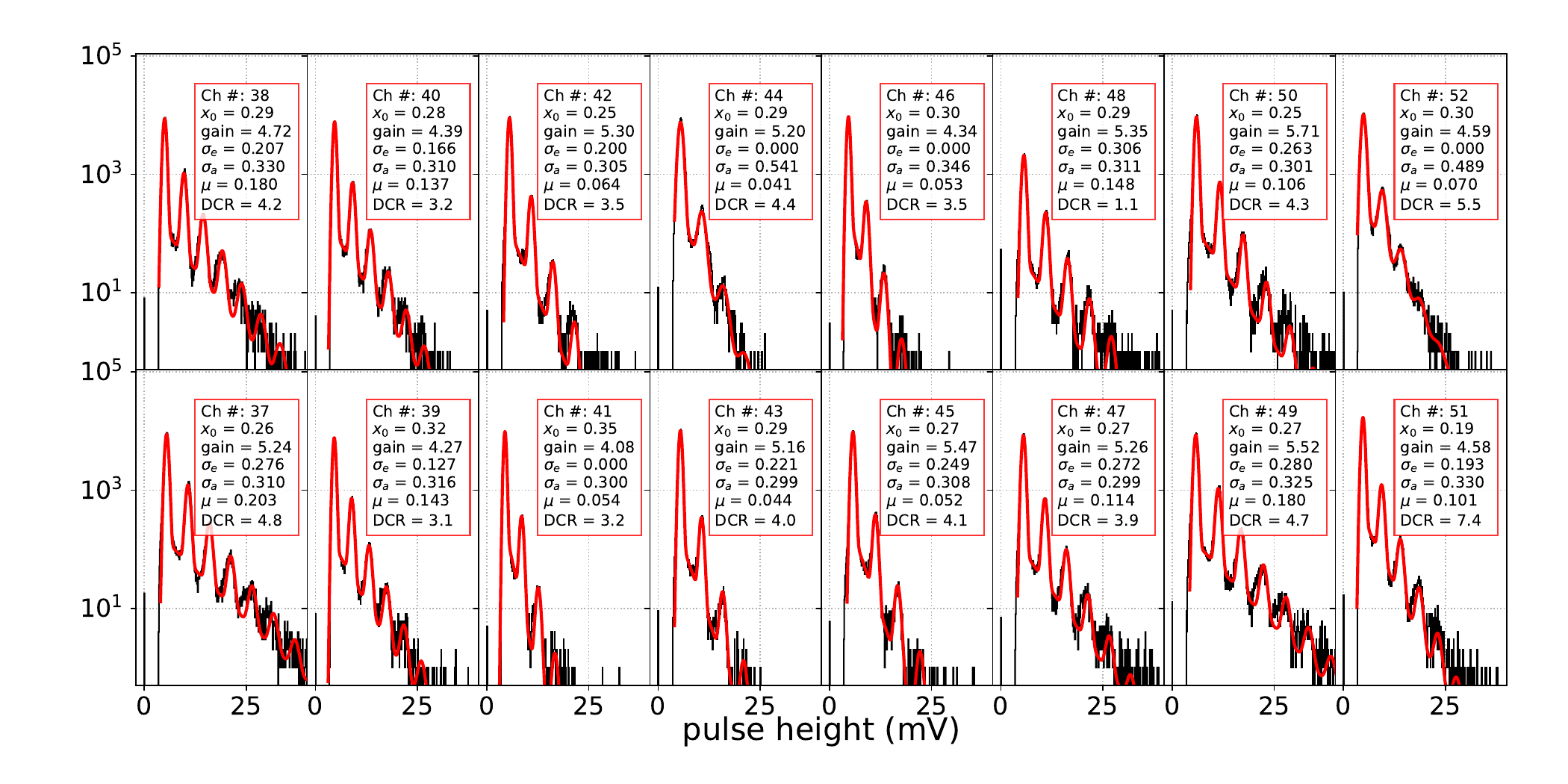}
\caption{\label{fig:dcrfit2} Fits to the pulse-height spectra measured from the dark count data-set for all channels on SCEMA board 2. The fitting model is described in \secref{sec:dcrintro}. The gain has units of filtered-mV per photon, the crosstalk parameter is unitless, and the dark count rate has units of MHz. Channel 48 is semi-functional and shows an artificially low DCR.  }
\end{figure}

\begin{figure}[htbp!]
\centering 
\includegraphics[width=.99\textwidth]{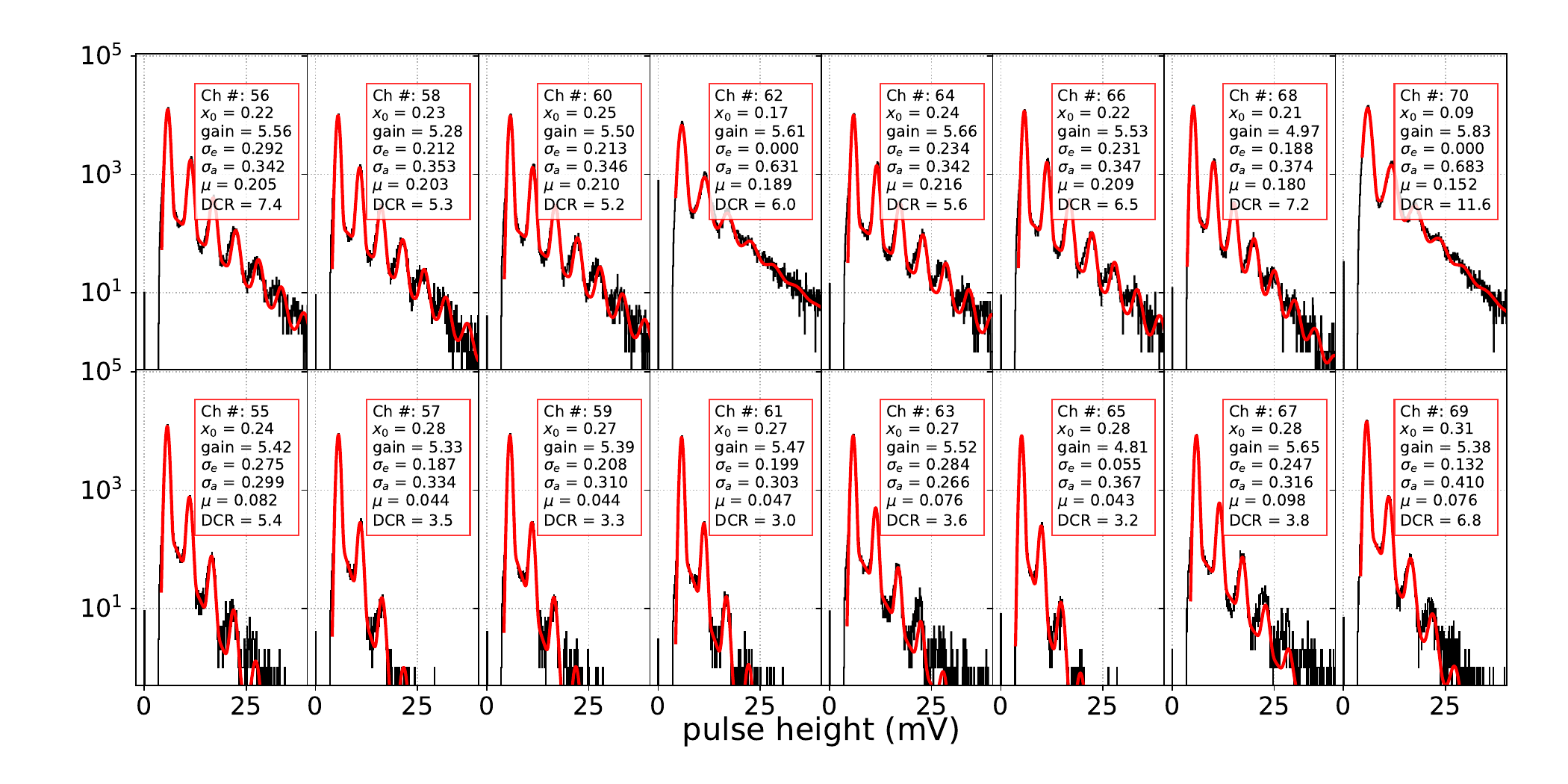}
\caption{\label{fig:dcrfit3} Fits to the pulse-height spectra measured from the dark count data-set for all channels on SCEMA board 3. The fitting model is described in \secref{sec:dcrintro}. The gain has units of filtered-mV per photon, the crosstalk parameter is unitless, and the dark count rate has units of MHz.  }
\end{figure}

\begin{figure}[htbp!]
\centering 
\includegraphics[width=.99\textwidth]{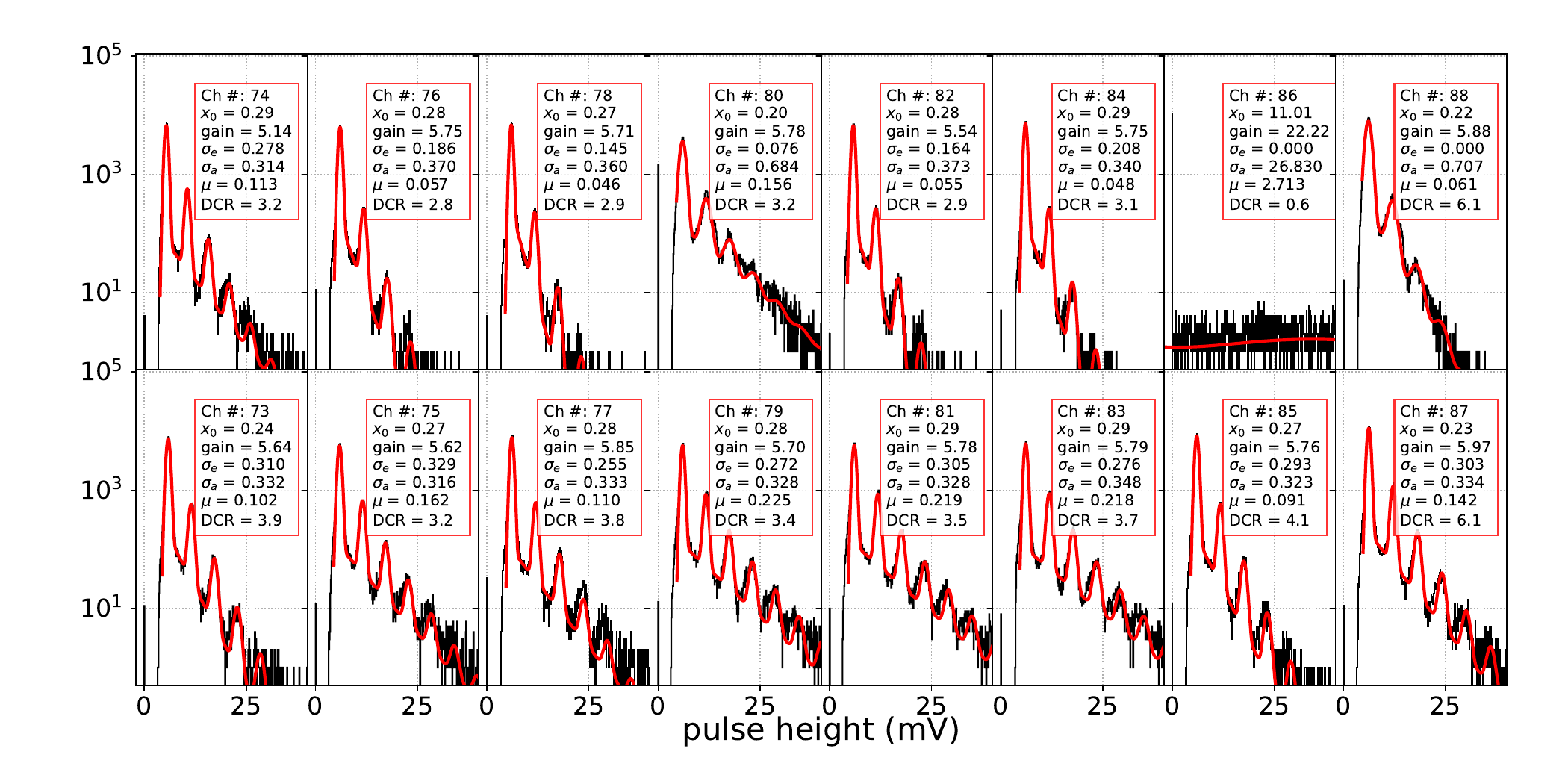}
\caption{\label{fig:dcrfit4} Fits to the pulse-height spectra measured from the dark count data-set for all channels on SCEMA board 4. The fitting model is described in \secref{sec:dcrintro}. The gain has units of filtered-mV per photon, the crosstalk parameter is unitless, and the dark count rate has units of MHz. Channel 86 is non-functional. }
\end{figure}

\begin{figure}[htbp!]
\centering 
\includegraphics[width=.99\textwidth]{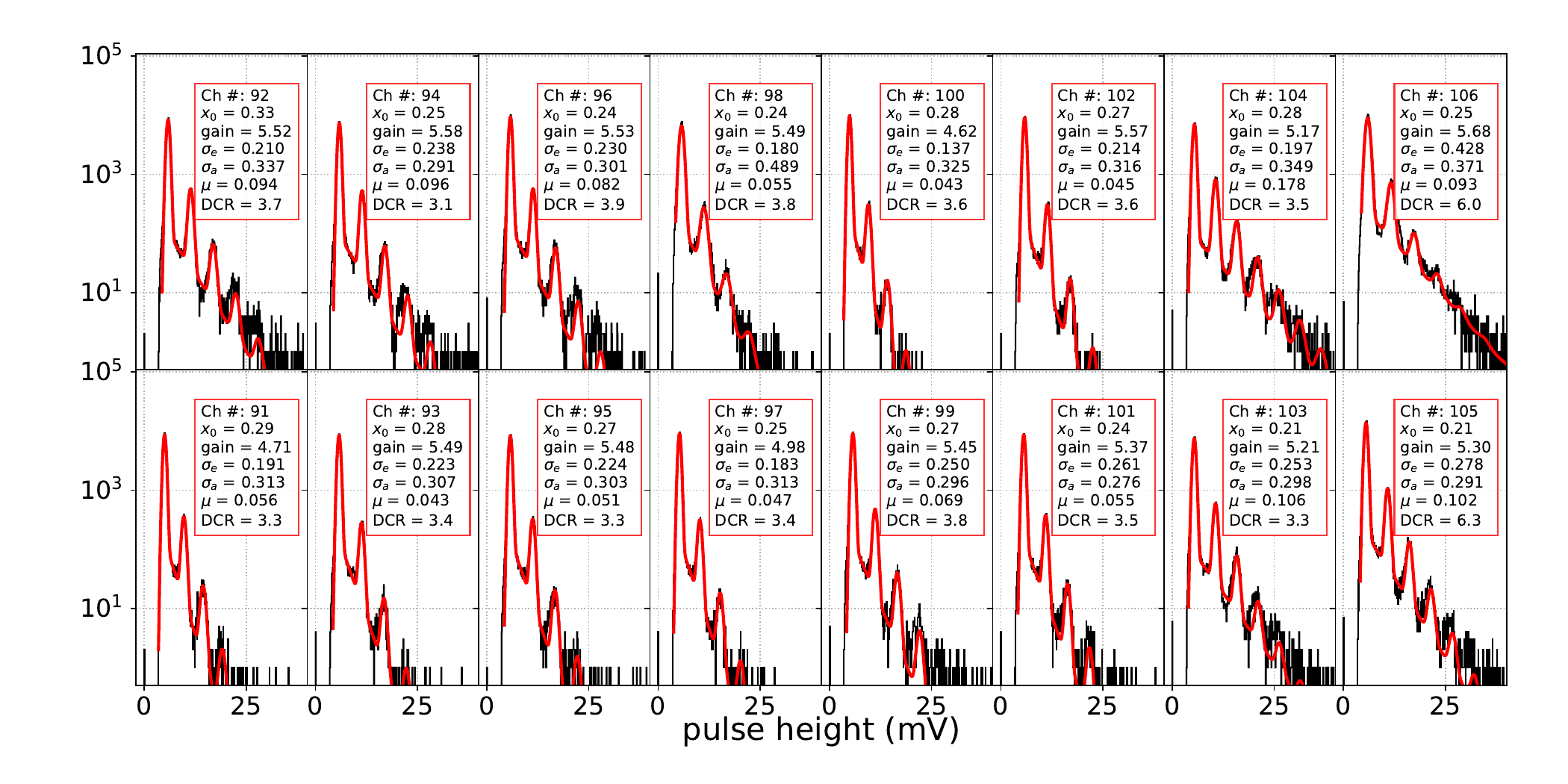}
\caption{\label{fig:dcrfit5} Fits to the pulse-height spectra measured from the dark count data-set for all channels on SCEMA board 5. The fitting model is described in \secref{sec:dcrintro}. The gain has units of filtered-mV per photon, the crosstalk parameter is unitless, and the dark count rate has units of MHz. }
\end{figure}

\begin{figure}[htbp!]
\centering 
\includegraphics[width=.99\textwidth]{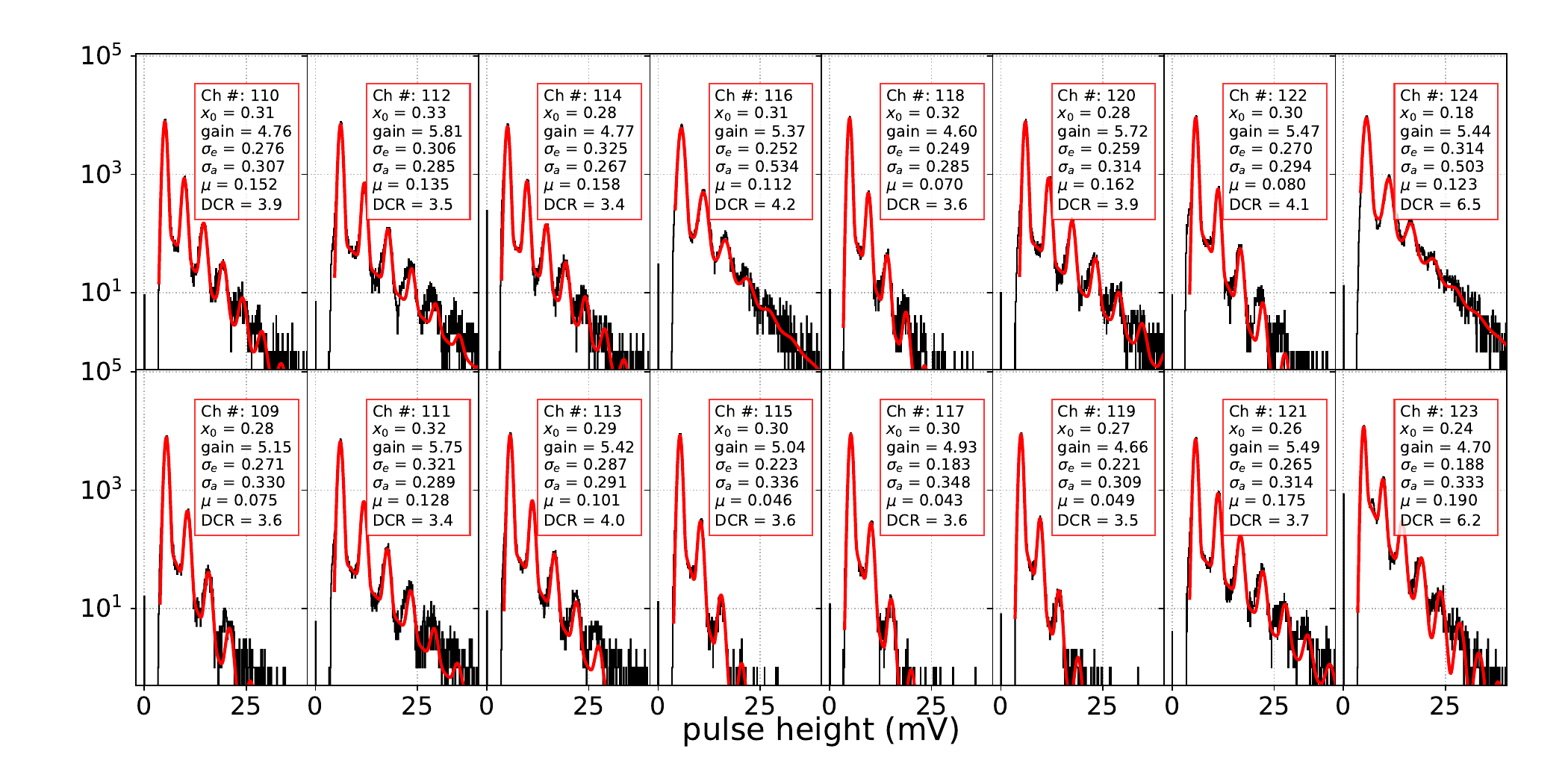}
\caption{\label{fig:dcrfit6} Fits to the pulse-height spectra measured from the dark count data-set for all channels on SCEMA board 6. The fitting model is described in \secref{sec:dcrintro}. The gain has units of filtered-mV per photon, the crosstalk parameter is unitless, and the dark count rate has units of MHz.  }
\end{figure}

\begin{figure}[htbp!]
\centering 
\includegraphics[width=.99\textwidth]{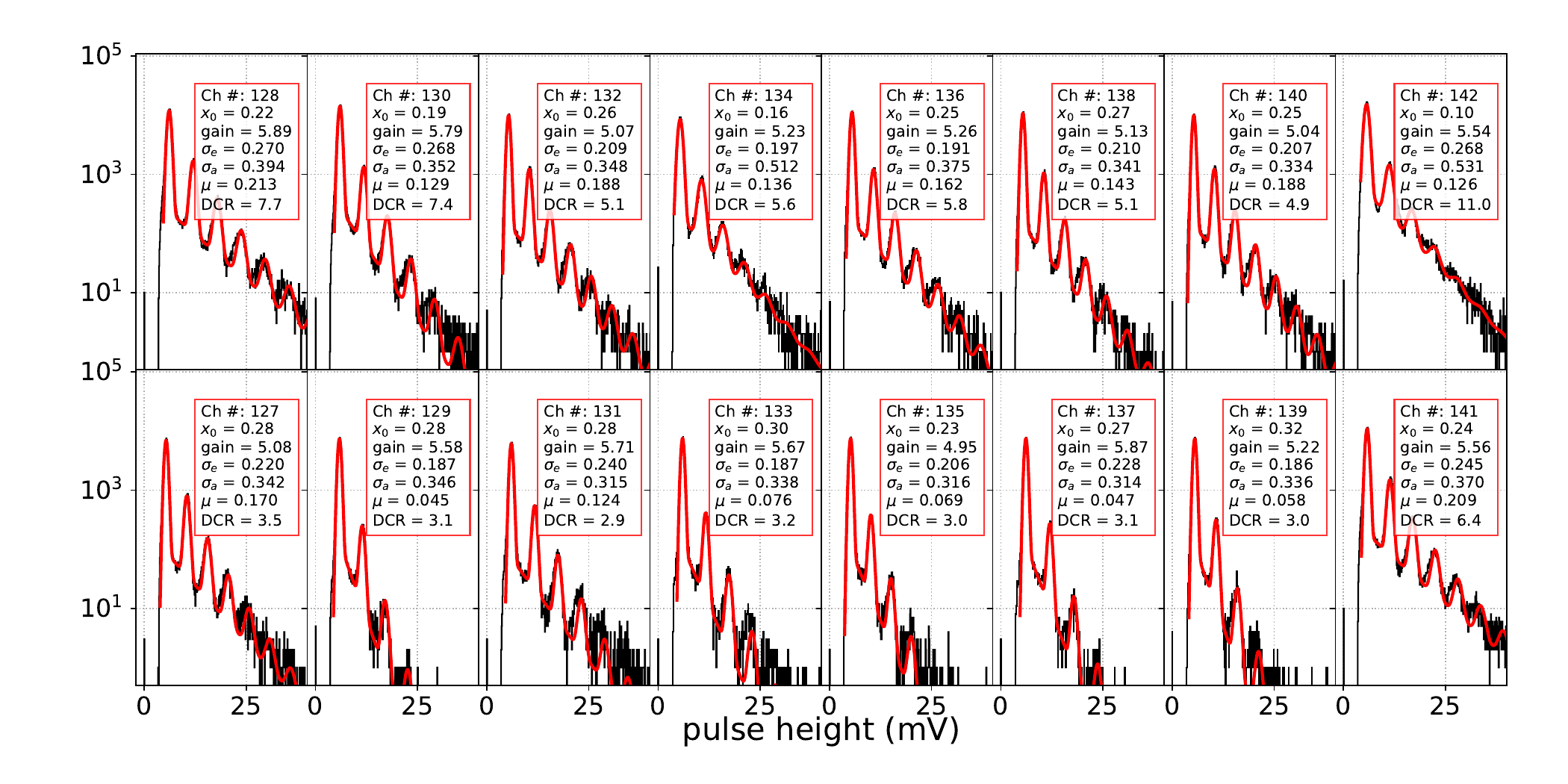}
\caption{\label{fig:dcrfit7} Fits to the pulse-height spectra measured from the dark count data-set for all channels on SCEMA board 7. The fitting model is described in \secref{sec:dcrintro}. The gain has units of filtered-mV per photon, the crosstalk parameter is unitless, and the dark count rate has units of MHz.   }
\end{figure}



\end{document}